\documentclass[12pt]{article}

\usepackage{graphicx}
\usepackage{amsmath}
\usepackage{amssymb}
\usepackage{multirow}
\usepackage{url}

\allowdisplaybreaks
\def\CR{\nonumber \\}
\def\eq#1{(\ref{#1})}
\def\[#1\]{\begin{align}#1\end{align}}

\begin{document}

\begin{titlepage}
\title{
\hfill\parbox{4cm}{ \normalsize YITP-16-99}\\ 
\vspace{1cm} 
Equation of motion of canonical tensor model\\ and Hamilton-Jacobi equation of general relativity}
\author{
Hua Chen$^a$\footnote{hua.chen@yukawa.kyoto-u.ac.jp}, 
Naoki Sasakura$^{a}$\footnote{sasakura@yukawa.kyoto-u.ac.jp}, 
Yuki Sato$^{b}$\footnote{Yuki.S@chula.ac.th}
\\
$^{a}${\small{\it Yukawa Institute for Theoretical Physics, Kyoto University,}}
\\ {\small{\it  Kitashirakawa, Sakyo-ku, Kyoto 606-8502, Japan,}}
\\
$^{b}${\small{\it Department of Physics, Faculty of Science, Chulalongkorn University}}
\\ {\small{\it Thanon Phayathai, Pathumwan, Bangkok 10330, Thailand.}}\\
}

\date{\today}
\maketitle
\thispagestyle{empty}
\begin{abstract}
\normalsize
The canonical tensor model (CTM) is a rank-three tensor model formulated as a totally constrained system
in the canonical formalism. 
The constraint algebra of CTM has a similar structure as that of the ADM formalism of general relativity,
and is studied as a discretized model for quantum gravity.
In this paper, we analyze the classical equation of motion (EOM) of CTM in a formal continuum limit
through a derivative expansion of the tensor of CTM up to the fourth order, and show that it is the same as the
EOM of a coupled system of gravity and a scalar field derived from the Hamilton-Jacobi equation
with an appropriate choice of an action.
The action contains a scalar field potential of an exponential form, and 
the system classically respects a dilatational symmetry. We find that the system has a critical dimension,
given by six, over which it becomes unstable due to the wrong sign of the scalar kinetic term.
In six dimensions, de Sitter spacetime becomes a solution to 
the EOM, signaling the emergence of a conformal symmetry, while the time evolution of 
the scale factor is power-law in dimensions below six.
\end{abstract}
\end{titlepage}

\section{Introduction}
The tensor model was first introduced in \cite{Ambjorn:1990ge,Sasakura:1990fs,Godfrey:1990dt} 
as an analytical description of simplicial quantum gravity in dimensions higher than 
two\footnote{See, however, \cite{Fukuma:2015xja,Fukuma:2015haa,Fukuma:2016zea} 
for a matrix-model-like approach to three-dimensional quantum gravity.} 
by generalizing the matrix model, which successfully describes the two dimensional case.
While the original tensor models are still remaining merely as formal descriptions due to some difficulties, 
the analyses of the more successful model, the colored tensor model 
\cite{Gurau:2009tw}, have produced various interesting analytical results concerning the simplicial quantum gravity
in dimensions higher than two \cite{Gurau:2011xp}. Among them, it has been shown that the dominant contributions 
of simplicial complexes generated 
from the colored tensor model are branched polymers \cite{Bonzom:2011zz,Gurau:2013cbh}. 
Since the structure of branched polymers is far from the classical spacetime picture of our universe, 
it seems difficult to consider the tensor model as a sensible model of quantum gravity, which
should produce wide and smooth spacetimes in certain classical regimes.   

On the other hand, while the models above basically concern the Euclidean case, 
it has been shown that Causal Dynamical Triangulation (CDT),
which is the simplicial quantum gravity with a causal structure, 
successfully produces the 3+1 dimensional world similar to our universe \cite{Ambjorn:2004qm},
while Dynamical Triangulation, which is the Euclidean version, does not\footnote{
When coupling many U$(1)$-fields, 
the authors in \cite{Horata:2000eg} found a promise of 
a phase transition higher than first order, 
which, however, is in conflict with the result in \cite{Ambjorn:1999ix}.
}.
The comparison between the two versions suggests that a causal structure is essentially 
important for the emergence of a classical spacetime in quantum gravity. 
This motivated one of the present authors to formulate
a rank-three tensor model as a totally constrained system in the canonical formalism,
which we call Canonical Tensor Model (CTM) \cite{Sasakura:2011sq,Sasakura:2012fb}.
The constraints of CTM are composed of kinematical symmetry generators and 
those analogous to the Hamiltonian constraint in the Arnowitt-Deser-Misner (ADM) formalism  
\cite{Arnowitt:1960es,Arnowitt:1962hi}, 
and form a first-class constraint algebra with a non-linear structure.  
In fact, the algebraic structure of the constraints is very similar to that of 
the ADM formalism of general relativity (GR), and 
it can be shown \cite{Sasakura:2015pxa} that, in a formal continuum limit, the constraint algebra of CTM
agrees with that of the ADM formalism of GR\footnote{As well, a certain minisuperspace model of GR can be derived from
CTM \cite{Sasakura:2014gia}.}.
This is of physical importance, since the algebraic closure of the ADM constraints assures
the spacetime covariance of locally defined time evolutions, which is an essence of GR \cite{Hojman:1976vp}.

The main purpose of this paper is to pursue this correspondence further.
We will analyze the classical equation of motion (EOM) of CTM in a formal continuum limit 
through a derivative expansion of the tensor of CTM up to the fourth order, 
and will show that it is the same as that of a coupled system of gravity and a scalar field derived from the Hamilton-Jacobi 
equation with an appropriate choice of an action.  
The action has an exponential potential of the scalar field, and the system is 
classically invariant under a dilatational symmetry. 
Interestingly, the action is meaningful only 
in spatial dimensions $2\leq d \leq 6$, and the system becomes unstable
in $d>6$ due to the wrong sign of the scalar kinetic term. In the critical dimension $d=6$, 
de Sitter spacetime becomes a solution to the EOM, signaling the emergence of a conformal symmetry.

The present work may also have some implications to renormalization-group (RG) flow equations of field theories. 
It has been argued \cite{Sasakura:2015xxa,Sasakura:2014zwa,Sasakura:2014yoa} 
that the Hamiltonian constraints of CTM generate the RG flows of statistical systems on random networks \cite{revnetwork},
which can equivalently be described by randomly connected tensor networks.
In addition, it has been shown \cite{Chen:2016xjx} that classical spaces emerge on boundaries of randomly 
connected tensor networks by appropriately choosing the tensors. Therefore, it can be expected that 
the Hamiltonian constraints would generate RG flows of effective field theories on such emergent spaces.
If so, the present work would give a hint to the connection between RG flows of field theories and gravity, 
which is indeed the subject of the so-called holographic RG (See \cite{Fukuma:2002sb} for a review.).

We heavily used a Mathematica package ``xTensor" \cite{xtensor} to perform tensorial computations in this paper.
The mathematica programs we used can be downloaded from one of the author's 
homepage \cite{sasahome}.

This paper is organized as follows. 
In Section~\ref{sec:review}, we review CTM. 
In Section~\ref{sec:repP}, we define the fields of CTM in a formal continuum limit in terms of a
derivative expansion of the tensor of CTM up to the fourth order. There are four fields, a rank-0,2,3,4 tensor
field with the weight of negative half-density.
In Section~\ref{sec:gauge}, we study the kinematical symmetry of CTM in the continuum limit. Up to the fourth order, 
we find two gauge symmetries, the diffeomorphism and a spin-three symmetry.
In Section~\ref{sec:fixed}, by deleting the rank-3 and rank-4 fields by the spin-three gauge symmetry and the EOM, respectively,
we write down the EOM of the remaining fields, the rank-0 and rank-2 fields,
in a static background geometry. 
In Section~\ref{sec:backmetric}, we discuss another gauge symmetry which allows us to freely transform the background metric. 
Then, in Section~\ref{sec:identify}, the background metric is gauge-fixed to a combination of the fields
so as to remove the odd situation that there exists a static spin-two field, the background metric, other than
the rank-2 field of CTM. The EOM with the gauge-fixing condition is written down. 
In Section~\ref{sec:delete}, we rewrite the EOM after deleting the weights of the fields. 
In Section~\ref{sec:reparametrization}, we 
perform a reparameterization of the fields so that there are no spatial derivative terms of the lapse function in EOM. 
This is the final form of the EOM of CTM, which is comparable with that of a gravitational system in field theory. 
In Section~\ref{sec:conttheory}, we show that the EOM of CTM can be made coincident with
that of a coupled system of gravity and a scalar field 
derived from the Hamilton-Jacobi equation by an appropriate choice of an action.
We find a critical spatial dimension of the gravitational system, given by six, 
over which the system becomes unstable due to the wrong sign of the kinetic term of the scalar field.
In Section~\ref{sec:mini}, we discuss the time evolution of the scale factor.  At the critical dimension, 
de Sitter spacetime is a solution to the EOM, signaling the emergence of a conformal symmetry, 
while the time evolution of the scale factor has a power-law behavior below the critical dimension.
Section~\ref{sec:summary} is devoted to the summary and future prospects. 
 
\section{Review of CTM}
\label{sec:review}
In this section we review the canonical tensor model (CTM) 
\cite{Sasakura:2011sq, Sasakura:2012fb}, 
explaining its current status. 

We consider a Hamiltonian system such that 
the dynamical variables are the real symmetric rank-three tensors, 
$M_{abc}$ and $P_{abc}$ $(a,b,c=1,2,\cdots, \mathcal{N})$, 
which are canonically conjugate 
in the sense that they satisfy the following Poisson bracket:
\[
\{ 
M_{abc}, P_{def}
\}
= 
\sum_{\sigma}
\delta_{a \sigma_d}\delta_{b \sigma_e}\delta_{c \sigma_f}, 
\ \ \ 
\{ 
M_{abc}, M_{def}
\}
= 
\{ 
P_{abc}, P_{def}
\}
=0,  
\label{eq:poissopn}
\]     
where the summation is over all the permutations of $d$, $e$ and $f$, 
reflecting the real symmetric nature of the tensors. 
Here, it would be natural to introduce the O($\mathcal{N}$) transformation
as a kinematical symmetry of the system,  
\[
\begin{split}
&M_{abc} \to M'_{abc} = L_{aa'}L_{bb'}L_{cc'}M_{a'b'c'}, \\
&P_{abc} \to P'_{abc} = L_{aa'}L_{bb'}L_{cc'}P_{a'b'c'},
\end{split}
\label{eq:ontransformation}
\]
where the repeated indices are summed over and $L$ is an O($\mathcal{N}$) matrix, 
since quantities constructed by the tensors 
with all indices being contracted are invariant under the O$(\mathcal{N})$ transformation. 
The Hamiltonian of CTM is given as follows:
\[
H_{CTM}
= n_a \mathcal{H}_a + n_{ab} \mathcal{J}_{ab},
\label{eq:ctmhamiltonian}
\] 
where $n_a$ and $n_{ab} (=-n_{ba})$ are non-dynamical Lagrange's multipliers, 
and
\[
&\mathcal{H}_a 
=\frac{1}{2} \left(P_{abc}P_{bde}M_{cde} - \lambda M_{abb} \right), 
\label{eq:ctmhamiltonianconstraint}
\\
&\mathcal{J}_{ab}
=-\mathcal{J}_{ba}
= \frac{1}{4} \left( P_{acd}M_{bcd} - P_{bcd}M_{acd} \right), 
\label{eq:ctmmomentumconstraint} 
\]
in which $\lambda$ is a cosntant. 
Imitating the nomenclatures in the Arnowitt-Deser-Misner (ADM) formalism of general relativity, 
$\mathcal{H}_a$ and $\mathcal{J}_{ab}$ are dubbed as 
Hamiltonian constraint and momentum constraint, respectively, 
and they form the following first-class constraint Poisson algebra: 
\[
\begin{split}
& \{ \mathcal{H} (\xi^1), \mathcal{H} (\xi^2) \} 
= \mathcal{J} ( [ \tilde{\xi}^1, \tilde{\xi}^2 ] + 2 \lambda\,  \xi^1 \wedge \xi^2 ), \\
& \{ \mathcal{J} (\eta), \mathcal{H} (\xi) \} 
= \mathcal{H} (\eta \xi), \\
&\{ \mathcal{J} (\eta^1), \mathcal{J} (\eta^2) \} 
= \mathcal{J} ([ \eta^1, \eta^2 ]), 
\label{eq:ctmconstraintalgebra}
\end{split}
\]
where $\mathcal{H}(\xi) := \xi_a \mathcal{H}_a$, $\mathcal{J}(\eta) := \eta_{ab}\mathcal{J}_{ab}$, 
and $\tilde{\xi}_{ab}:= P_{abc}\xi_c$. 
In (\ref{eq:ctmconstraintalgebra}), the bracket $[\ , \ ]$ denotes the matrix commutator, 
and $(\xi^1\wedge \xi^2)_{ab}:= \xi^1_a \xi^2_b - \xi^2_a \xi^1_b$. 
One notices that $\mathcal{J}$ serves as the generators of SO($\mathcal{N}$),
infinitesimally representing the kinematical symmetry of the system.
The form of the Hamiltonian constraint has been uniquely fixed 
by the following five assumptions: 
the Hamiltonian constraint  
(I) carries only one index, 
(II) forms a closed Poisson algebra with $\mathcal{J}$, 
(III) preserves the time reversal symmetry, $M_{abc} \to M_{abc}$ and $P_{abc}\to - P_{abc}$, 
(IV) consists of terms cubic at most, 
and 
(V) allows only ``connected terms,'' 
e.g., 
$P_{abc}P_{bde}M_{cde}$ is allowed but $M_{abb}P_{cde}P_{cde}$ is not allowed \cite{Sasakura:2012fb}.
With the closed Poisson algebra (\ref{eq:ctmconstraintalgebra}) of the constraints, CTM is a totally constrained system 
governed by the Hamiltonian (\ref{eq:ctmhamiltonian}). 
The interest of this paper is the classical equation of motion (EOM) of $P$ with $\lambda=0$, which 
is given by
\[
\begin{split}
\frac{\text{d}}{\text{d}t} P_{abc} 
= \{P_{abc}, H_{CTM}^{\lambda=0} \} 
= - \frac{1}{2} 
\sum_{\sigma}
\left(
n_d P_{de \sigma_{a}} P_{\sigma_{b}\sigma_{c}e} 
+  n_{d \sigma_{a}} P_{\sigma_{b}\sigma_{c}d}
\right). 
\label{eq:ctmeom}  
\end{split}
\]
The variable $M_{abc}$ will play no roles in this paper.

Quite remarkably, CTM is closely related to general relativity in arbitrary dimensions 
in the following sense. 
Firstly, for $\mathcal{N}=1$ case, the Hamiltonian (\ref{eq:ctmhamiltonian}) agrees with that of a certain minisuperspace 
model of GR in arbitrary dimensions, 
if we consider 
the modulus of the tensor, $|M_{111}|$, is proportional to the spatial volume in the minisuperspace model \cite{Sasakura:2014gia}. 
Secondly, in a formal continuum limit with $\mathcal{N}\to \infty$, 
the Poisson algebra (\ref{eq:ctmconstraintalgebra}) coincides with the Dirac algebra in the ADM formalism \cite{Sasakura:2015pxa}. 
In this paper we take this argument one step further: 
we will analyze the EOM (\ref{eq:ctmeom}) of CTM in a formal continuum limit
through a derivative expansion of $P$ up to the fourth order, and will
show that it agrees with the EOM of a coupled system of gravity and a scalar field 
derived from the Hamilton-Jacobi equation with an appropriate choice of an action.

\section{Representation of the tensor in a derivative expansion}
\label{sec:repP}
In this paper, we consider CTM in a formal continuum limit.
We leave aside for future study the question of dynamics why CTM can be studied in the continuum manner: we 
simply assume that there exist some regimes where the continuum description is valid.  
The basic strategy to treat CTM in this limit is the same 
as that in the previous papers \cite{Sasakura:2015pxa,Chen:2016xjx}. 
We formally replace the discrete values of the indices to the $d$-dimensional spatial coordinates: 
\[
a \rightarrow x \in R^d.
\] 
Namely, the tensor $P_{xyz}$ is a function of three $d$-dimensional coordinates $x,y,z$, symmetric under arbitrary permutations.
We further assume a locality: $P_{xyz}$ takes non-vanishing values, only when $x,y,z$ are 
in the neighborhood, $x\sim y \sim z$. Mathematically, this can be formulated by that $P_{xyz}$ is a 
distribution described by delta functions and their 
derivatives\footnote{In \cite{Sasakura:2015pxa}, the mathematical formulation is presented
differently as a moment expansion in coordinates. 
Though they are essentially the same from the physical point of view, the present formulation in terms of distributions is superior to 
the former one in the sense that the covariance can be easily incorporated. }:
 $P_{xyz}\sim \delta^d(x-y)\delta^d (y-z)+\hbox{derivatives of }\delta^d(x-y)\delta^d (y-z)$. 
We also assume that we can terminate the derivative expansion at a certain order. From the 
physical point of view, this is an assumption that the scale of the physical process of our interest is 
much larger than the fuzziness of the locality of the space. 
In general, it is more convenient to use test functions to describe distributions 
rather than directly dealing with $\delta$-functional expressions. 
So, let us consider a contraction of $P$ with a test function $f$ up to the fourth order of derivatives as follows:
\[
\begin{split}
P f^3&:=
\int d^dx d^dy d^dz\, P_{xyz} f(x)f(y)f(z) \\
&=\int d^dx \left( \beta f^3 +  \beta^{\mu\nu}f^2 f_{,\mu\nu}+ \beta^{\mu\nu\rho} f^2 f_{,\mu\nu\rho}
+\beta^{\mu\nu,\rho\sigma}f f_{,\mu\nu}f_{,\rho\sigma}+{\cal O}(\nabla^5)\right),
\end{split}
\label{eq:pf3}
\]
where, for brevity, the arguments $x$ of $\beta$'s and $f$ are suppressed in the last line, 
and the greek indices represent spatial directions, e.g., $\mu=1,2,\cdots,d$.
Here, the test function $f$ is assumed to have a compact support, 
and the indices of $f$ represent the covariant derivatives associated with a background metric $g_{\mu\nu}$, 
i.e., $f_{,\mu\nu}:=\nabla_\mu \nabla_{\nu} f,\ f_{,\mu\nu\rho}:=\nabla_\mu \nabla_{\nu} \nabla_\rho f$.
As will be explained in more detail in Section~\ref{sec:gauge}, the test function is not a scalar, but 
must be treated as a scalar half-density. Therefore, the covariant derivatives are defined with 
a weight contribution: $\nabla_\mu f=(\partial_\mu -\frac{1}{2} \Gamma_\mu)f$ with 
$\Gamma_{\mu}:=\Gamma_{\mu\nu}^\nu$,
$\nabla_\mu \nabla_\nu f=(\partial_\mu -\frac{1}{2}\Gamma_\mu)\nabla_\nu f -\Gamma_{\mu\nu}^\rho \nabla_\rho f$, and so on. The tensor fields, $\beta^{\mu\nu}$ and $\beta^{\mu\nu\rho}$, are symmetric, 
and the field $\beta^{\mu\nu,\rho\sigma}$ has the pairwise symmetries,
\[
\beta^{\mu\nu,\rho\sigma}=\beta^{\nu\mu,\rho\sigma}=\beta^{\mu\nu,\sigma\rho}=
\beta^{\rho\sigma,\mu\nu}.
\]
Thus, up to the fourth order, the ``components" of $P$ are represented by the four fields, 
$\beta(x),\beta^{\mu\nu}(x),\beta^{\mu\nu\rho}(x)$, and $\beta^{\mu\nu,\rho\sigma}(x)$.
Because of the weight of $f$ and the invariance of $Pf^3$, 
these fields are assumed to have the weight of negative half-density
(the details will be given in Section~\ref{sec:gauge}):
\[
\begin{split}
&[f]=\frac{1}{2}, \\
&[\beta]=[\beta^{\mu\nu}]=[\beta^{\mu\nu\rho}]=[\beta^{\mu\nu,\rho\sigma}]=-\frac{1}{2}.
\end{split}
\label{eq:weightbeta}
\]
Here, $[X]$ denotes the weight of a quantity $X$, meaning that $X$ has the same weight as $g^\frac{[X]}{2}$
($g:=\hbox{Det}[g_{\mu\nu}]$). These weights cancel the weight of the integration measure $d^dx$ to secure the 
invariance of $Pf^3$.  

Here, we will explain more details about the derivative expansion \eq{eq:pf3}. 
Firstly, as proven in Appendix~\ref{app:pf3}, a totally symmetric rank-three tensor 
can be fully characterized by the values of the contraction with an arbitrary vector $\phi$\,: 
$P_{abc}\phi_a \phi_b \phi_c \hbox{ for }^\forall \phi$.
Thus, it is enough to know $Pf^3$ for arbitrary $f$ as in \eq{eq:pf3} for the full characterization of $P$, instead of 
considering three different functions for the three indices. 
Secondly, throughout this paper, we will consider the derivative expansion of $P$ up to the fourth order of derivatives, 
as in \eq{eq:pf3}. 
The reason is that we are interested in the equations of motion (EOM) of the fields $\beta,\beta^{\mu\nu}$ up to the 
second order of derivatives: as will be discussed later, these fields describe a coupled system of gravity and a scalar field,
which is of physical interest. To correctly describe the EOM of 
$\beta^{\mu\nu}$ (and $\beta$) up to the second derivatives, 
it is necessary to include the fourth order of derivatives in the expansion of $P$ as in \eq{eq:pf3}.
As one can prove, an independent set of fields describing $P$ up to the fourth order
are exhausted by the set shown in \eq{eq:pf3}. 
More details are given in Appendix~\ref{app:fourth} and \ref{app:derivative}.
Lastly, we have introduced a background metric $g_{\mu\nu}$, which can be taken arbitrary. 
As will be explained in detail in Section~\ref{sec:backmetric},
the introduction of the background metric does not change the physical contents,
but simply redefines the fields with a linear recombination of them.
In fact, we will see that there exists a gauge symmetry which allows one to freely change the background metric
with simultaneous change of the fields, and will ultimately
gauge-fix the background metric to a certain combination of the fields.

In the analysis of the EOM \eq{eq:ctmeom} of CTM, 
it is necessary to have an expression corresponding to $3 P_{abc}\phi_b \phi_c$. 
In the continuum limit, one can obtain this by the functional derivative of $Pf^3$ in \eq{eq:pf3}:
\[
\begin{split}
P[f,f]:=&\frac{\delta}{\delta f(x)} Pf^3 \\
=&3 \beta f^2 + 2 \beta^{\mu\nu} f f_{,\mu\nu} + (\beta^{\mu\nu} f^2 )_{,\mu\nu} +2 \beta^{\mu\nu\rho} f f_{,\mu\nu\rho}
-(\beta^{\mu\nu\rho}f^2)_{,\mu\nu\rho} +\beta^{\mu\nu,\rho\sigma} f_{,\mu\nu} f_{,\rho\sigma} \\
&+2 (\beta^{\mu\nu,\rho\sigma} f f_{,\mu\nu})_{,\rho\sigma} +{\cal O}(\nabla^5)\\
=&(3 \beta +\beta^{\mu\nu}_{,\mu\nu}-\beta^{\mu\nu\rho}_{,\mu\nu\rho})f^2+(4 \beta^{\mu\nu}_{,\nu}
-6 \beta_{,\nu\rho}^{\mu\nu\rho}) f f_{,\mu}+(2 \beta^{\mu\nu}-6 \beta^{\mu\nu\rho}_{,\rho}) f_{,\mu} f_{,\nu} \\
&+(4 \beta^{\mu\nu}-6 \beta_{,\rho}^{\mu\nu\rho}+
2\beta^{\mu\nu,\rho\sigma}_{,\rho\sigma}) f f_{,\mu\nu}  
+(-6 \beta^{\mu\nu\rho}+4\beta^{\mu\sigma,\nu\rho}_{,\sigma})
f_{,\mu} f_{,\nu\rho} \\
&+4\beta^{\mu\nu,\rho\sigma}_{,\sigma} f f_{,\rho\mu\nu} 
+3\beta^{\mu\nu,\rho\sigma}f_{,\mu\nu}f_{,\rho\sigma}
+4\beta^{\mu\nu,\rho\sigma}f_{,\mu} f_{,\nu\rho\sigma}
+2\beta^{\mu\nu,\rho\sigma}f f_{,\mu\nu\rho\sigma}+{\cal O}(\nabla^5).
\end{split}
\label{eq:pff}
\]

Similarly, one can define an expression corresponding to $3 P_{abc}\phi^1_a \phi_b^2$ for two different vectors $\phi^{1,2}$. 
This is denoted by $P[f,g]$, and is defined by an obvious generalization: 
putting $f,g$ into two $f$'s of each term on the right-hand side of \eq{eq:pff}, and symmetrizing them.

\section{Kinematical symmetry in the continuum limit}
\label{sec:gauge}
CTM has the kinematical symmetry generated by the orthogonal group generators ${\cal J}_{ab}$. 
In the continuum limit, since the indices represent coordinates, 
${\cal J}_{xy}$ will become generators of local gauge transformations. 
In the derivative expansion, the gauge transformations are parameterized by 
tensor fields, like those in \eq{eq:pf3} for $P$.
Up to the fourth order, we will find two gauge transformations,
which are the diffeomorphism and a spin-three gauge transformation.

The orthogonal group transformation of CTM can be characterized by a linear transformation of $f_a$ which preserves
the norm square $f_a f_a$. In the continuum limit, this condition is translated to the invariance of 
\[
\Vert f \Vert^2\equiv \int d^dx\  f(x)f(x),
\label{eq:norm}
\]  
where $f(x)$ is considered to be a scalar half-density, and is assumed to have a compact support.
It is easy to show that \eq{eq:norm} is invariant under the following infinitesimal linear transformations,
\[
\begin{split}
\delta_1 f(x)&=\frac{1}{2} 
\left[\nabla_\mu( v^\mu(x) f(x)) + v^\mu(x) \nabla_\mu f(x)\right]=\frac{1}{2} v^\mu_{,\mu}(x) f(x)+ v^\mu(x) f_{,\mu}(x),\\
\delta_3 f(x)&=\frac{1}{2} \left[ \nabla_\mu\nabla_\nu\nabla_\rho( v^{\mu\nu\rho}(x) f(x)) + v^{\mu\nu\rho}(x) \nabla_\mu\nabla_\nu
\nabla_\rho f(x)\right] \\
&=\frac{1}{2} v^{\mu\nu\rho}_{,\mu\nu\rho}(x) f(x)+\frac{3}{2} v^{\mu\nu\rho}_{,\mu\nu}(x) f_{,\rho}(x)+\frac{3}{2} v^{\mu\nu\rho}_{,\mu}(x) f_{,\nu\rho}(x)
+ v^{\mu\nu\rho}(x) f_{,\mu\nu\rho}(x),
\end{split}
\label{eq:delta13}
\]
where $v^\mu$ and $v^{\mu\nu\rho}$ are a vector field and a symmetric rank-three tensor field, respectively, and 
$\nabla_\mu$ is the covariant derivative ($\nabla_\mu f=(\partial_\mu -\frac{1}{2} \Gamma_\mu)f$ with 
$\Gamma_\mu\equiv \Gamma_{\mu\nu}^\nu$, etc.).
Here we use the same simplified notations as in Section~\ref{sec:repP}, such as $f_{,\mu\nu}=\nabla_\mu\nabla_\nu f$.
Indeed,
\[
\delta_1 \Vert f \Vert^2 =2 \int d^dx\,  f(x) \delta_1 f(x)=\int d^dx \, f(x) \left[ \nabla_\mu( v^\mu(x) f(x)) 
+ v^\mu(x) \nabla_\mu f(x)\right]=0,
\]
because the integrand is a total derivative.\footnote{Note that $\Gamma$'s cancel out as 
$\nabla_\mu (v^\mu f^2)=(\partial_\mu +\Gamma_{\nu\mu}^\nu-\Gamma_\mu) (v^\mu f^2)=\partial_\mu (v^\mu f^2)$}
The invariance under $\delta_3$ can also be shown similarly by using partial integrations.
As can be seen in \eq{eq:delta13}, the transformation $\delta_1$ 
represents a diffeomorphism transformation, which transforms $f(x)$ as a scalar-half density, and 
$\delta_3$ represents a spin-three transformation.

Some comments are in order. Firstly, 
$v^{\mu\nu\rho}$ must be assumed to be symmetric to remove redundancies.
The reason is basically the same as that for the symmetry of $\beta$'s in \eq{eq:pf3}, which is explained 
in detail in Appendix~\ref{app:fourth}.
The anti-symmetric part of $f_{,\mu\nu\rho}$ in \eq{eq:delta13} can be rewritten 
in terms of the first derivative of $f$ by using the curvature tensor,
and therefore the anti-symmetric components of $v^{\mu\nu\rho}$ can be absorbed into $v_\mu$. 
Another comment is that one may consider a spin-two transformation with $v^{\mu\nu}$
in a similar manner. However, this is also redundant. 
The invariance of the norm \eq{eq:norm} requires that the transformation should be
in the form, $\delta_2 f=\nabla_\mu\nabla_\nu( v^{\mu\nu} f) - v^{\mu\nu} \nabla_\mu\nabla_\nu f$, with a minus
relative sign in this case. Then, the terms with the second derivative of $f$ cancel, and the transformation 
is equivalent to a diffeomorphism transformation with $v^\mu=v^{\mu\nu}_{,\nu}$. 
Finally, it is obvious that there exist an infinite tower of spin-odd transformations
which preserve \eq{eq:norm}. However, 
the transformations higher than spin-three 
are irrelevant in our treatment up to the fourth order of derivatives.
 
Let us define the transformations of $\beta$'s in \eq{eq:pf3} under $\delta_{1}$ and $\delta_{3}$,
by transferring the transformations of $f$ to $\beta$'s.
As for $\delta_1$, we obtain
\[
\begin{split}
\delta_1 \left(Pf^3\right)&=\int d^dx \left[ 
3\beta  f^2 \delta_1 f 
+\beta^{\mu\nu}\left( 2 f (\delta_1 f) f_{,\mu\nu} + f^2 (\delta_1 f)_{,\mu\nu} \right) \right. \\
&\hspace{6cm}
\left. +\beta^{\mu\nu,\rho\sigma}\left( (\delta_1f) f_{,\mu\nu} f_{,\rho\sigma}+2 f f_{,\mu\nu}(\delta_1 f)_{,\rho\sigma}\right)+{\cal O}(\nabla^5)\right]\\
&=\int d^dx \left[
(\delta_1\beta) f^3 + (\delta_1 \beta^{\mu\nu}) f^2 f_{,\mu\nu} 
+ (\delta_1 \beta^{\mu\nu\rho}) f^2 f_{,\mu\nu\rho}
+ (\delta_1 \beta^{\mu\nu,\rho\sigma}) f f_{,\mu\nu} f_{,\rho\sigma}+{\cal O}(\nabla^5)
\right],
\end{split}
\label{eq:derdel1}
\]
where
\[
\begin{split}
&\delta_1 \beta=- v^\mu \beta_{,\mu} + \frac{1}{2} v^\mu_{,\mu}\beta +{\cal O}(\nabla^3 ), \\
&\delta_1 \beta^{\mu\nu} =- v^\rho \beta^{\mu\nu}_{,\rho}+\frac{1}{2}v^\rho_{,\rho} \beta^{\mu\nu}+ v^\mu_{,\rho} \beta^{\rho\nu}+
v^\nu_{,\rho} \beta^{\mu\rho}+{\cal O}(\nabla^3 ), \\
&\delta_1 \beta^{\mu\nu\rho}={\cal O}(\nabla^2 ),\\
&\delta_1 \beta^{\mu\nu,\rho\sigma} ={\cal O}(\nabla ).
\end{split}
\label{eq:diffeo}
\]
To derive the result, we have performed some partial integrations to transform the first line of 
\eq{eq:derdel1} into the form of \eq{eq:pf3} in the second line.
We have assumed $\beta^{\mu\nu\rho}=0$ initially, which will be discussed later as a gauge condition
for the spin-three gauge symmetry.
The terms with ${\cal O}(\nabla^3)$ in $\beta$ and $\beta^{\mu\nu}$
can also be ignored, because our interest is up to the second derivatives for these fields. 
$\delta_1 \beta^{\mu\nu\rho}$ and $\delta_1 \beta^{\mu\nu,\rho\sigma}$ 
can be ignored, because they are of the fifth order of derivatives in \eq{eq:derdel1}.
 The result \eq{eq:diffeo} shows that $\beta$ transforms as a scalar of negative half-density, 
and $\beta^{\mu\nu}$ as a two-tensor of negative half-density.
Indeed, 
this coincides with the weight assignments \eq{eq:weightbeta},
what is apparently expected from the invariance of \eq{eq:pf3} under the diffeomorphism.

As for $\delta_3$,  in a similar manner, we obtain 
\[
\delta_3 (Pf^3)&=\int d^dx \left[ 3 \beta f^2 \delta_3 f + {\cal O}(\nabla^5 ) \right] \CR
&= \int d^dx \left[ (\delta_3 \beta) f^3 
+(\delta_3 \beta^{\mu\nu}) f^2 f_{,\mu\nu}+(\delta_3 \beta^{\mu\nu\rho}) f^2 f_{,\mu\nu\rho} + 
(\delta_3 \beta^{\mu\nu,\rho\sigma})  f f_{,\mu\nu} f_{,\rho\sigma}+{\cal O}(\nabla^5 )\right],
\]
where 
\[
\begin{split}
&\delta_3\beta = {\cal O}(\nabla^3 ),\\
&\delta_3 \beta^{\mu\nu}=\frac{9}{2}\beta v_{,\rho}^{\mu\nu\rho}, \\
&\delta_3 \beta^{\mu\nu\rho} =3 \beta v^{\mu\nu\rho},\\
&\delta_3 \beta^{\mu\nu,\rho\sigma} = {\cal O}(\nabla ).
\end{split}
\label{eq:delta3}
\]

The equation of motion \eq{eq:ctmeom} of CTM contains the second term
$\sum_\sigma n_{d \sigma_{a}} P_{\sigma_{b}\sigma_{c}d}$, which 
represents the freedom to perform the infinitesimal kinematical transformation along time evolution
by freely choosing $n_{ab}$ dependent on time. 
Within our approximation of the continuum limit, the transformations which are relevant
are $\delta_1$ and $\delta_3$. 
Thus, we can write \eq{eq:ctmeom} in a schematic manner as 
\[
\begin{split}
&\frac{d}{dt} \beta =(nPP)+\delta_1 \beta, \\
&\frac{d}{dt} \beta^{\mu\nu}=(nPP)^{\mu\nu}+\frac{9}{2}\beta v_{,\rho}^{\mu\nu\rho}+\delta_1 \beta^{\mu\nu}, \\ 
&\frac{d}{dt} \beta^{\mu\nu\rho}=(nPP)^{\mu\nu\rho}+3 \beta v^{\mu\nu\rho},\\
&\frac{d}{dt} \beta^{\mu\nu,\rho\sigma}=(nPP)^{\mu\nu,\rho\sigma},\\
\end{split}
\label{eq:schemEOM}
\]
where  we have used \eq{eq:diffeo} and \eq{eq:delta3}, and $(nPP),(nPP)^{\mu\nu},(nPP)^{\mu\nu\rho},
(nPP)^{\mu\nu,\rho\sigma}$ denote 
the spin-0,2,3,4 components of $\sum_{\sigma} n_{d} P_{\sigma_a de}P_{e\sigma_b\sigma_c}$, respectively.
Since $\delta_1$ describes the diffeomorphism,
the terms with $\delta_1$ in \eq{eq:schemEOM} correspond to the freedom to choose the shift-vector in the time-evolution 
in the ADM formalism of general relativity.
As for the spin-3 transformation, 
by setting $v^{\mu\nu\rho}=-(nPP)^{\mu\nu\rho}/3\beta$ under the assumption $\beta\neq0$,
we can make a tuning $\frac{d}{dt} \beta^{\mu\nu\rho}=0$.
In this manner, one can keep the gauge condition $\beta^{\mu\nu\rho}=0$, which gauges 
away the spin-3 component. As seen in \eq{eq:schemEOM}, 
by doing this gauge fixing, the time evolution of the spin-2 component will get a contribution by an amount,
\[
-\frac{3}{2}\beta \left(\frac{(nPP)^{\mu\nu\rho}}{\beta} \right)_{,\rho},
\]
from the infinitesimal spin-3 transformation.
Note that, even if $P$ has no spin-3 component, i.e. $\beta^{\mu\nu\rho}=0$, $(nPP)^{\mu\nu\rho}$ does not
vanish in general (This will be seen explicitly later.), 
and the spin-3 infinitesimal transformation must be carried out as above to keep $\beta^{\mu\nu\rho}=0$
along time evolution.
In later sections, this and similar procedures will frequently be used to remove the appearance of the spin-3 component.
In fact, the spin-3 component can appear not only from the right-hand side of the equation of motion \eq{eq:ctmeom}, 
but also from the left-hand side $\frac{d}{dt}P$, when the background metric has time-dependence as will be discussed in 
Section~\ref{sec:identify}.
This can also be removed by balancing it with the spin-3 transformation on the right-hand side
in a similar manner as above. 

\section{Equation of motion of CTM in a static background}
\label{sec:fixed}
In this section, we will study the continuum limit of the equation of motion (EOM) \eq{eq:ctmeom} of CTM in the case that
the background metric $g_{\mu\nu}$ is static. Let us take the contractions of both sides of \eq{eq:ctmeom} 
with a test function $f$ satisfying $\dot f=0$.
The left-hand side, $\frac{d}{dt} (Pf^3)$,  is simply given by \eq{eq:pf3} with $\beta$'s replaced by $\dot \beta$'s.
The right-hand side is given by   
\[
\delta Pf^3:=  \int d^dx\, n P[f,P[f,f]],
\label{eq:delpf3}
\]
where we have left aside the SO($\mathcal{N}$) rotational part of \eq{eq:ctmeom} 
for later discussions,  
have performed a replacement $n_a\rightarrow n(x)$, and an overall numerical factor has been 
absorbed into a constant rescaling of $n(x)$. By rewriting \eq{eq:delpf3} in the form of \eq{eq:pf3}, namely,
\[
\delta P f^3 = \int d^d x \left[
(\delta\beta)f^3+(\delta\beta^{\mu\nu} )f^2 f_{,\mu\nu}+(\delta\beta^{\mu\nu\rho} )f^2 f_{,\mu\nu\rho}+(\delta \beta^{\mu\nu,\rho\sigma})
f f_{,\mu\nu} f_{,\rho\sigma} +{\cal O}(\nabla^5)
\right],
\label{eq:canpf3}
\]
one can obtain the explicit expression of the right-hand side of the EOM for the fields $\beta$'s.
Here, note that a spin-three component $\delta \beta^{\mu\nu\rho}$ of $\delta P$ may appear
in general, even though the gauge condition $\beta^{\mu\nu\rho}=0$ is initially assumed on $P$.  

The symmetric two-tensor field $\beta^{\mu\nu}$ is particularly interesting from 
the view point of gravity.
The lowest order set of fields containing it is given by $\beta$ and $\beta^{\mu\nu}$.
Therefore, we want to compute $\delta \beta$ and $\delta \beta^{\mu\nu}$ up to the second order of derivatives,
which would be the minimum for physically interesting dynamics to be expected.
The wanted order about the latter field requires that our computations must be correct up to 
the fourth order in \eq{eq:canpf3}.
This means that $\delta \beta^{\mu\nu\rho}$ and $\delta \beta^{\mu\nu,\rho\sigma}$ must be 
computed up to the first and the zeroth order of derivatives, respectively.

It would seem that the fourth order terms\footnote{There exist no third order terms.} 
in $\delta \beta$ must also be included for 
the consistency of the fourth order computations. However,  the order of derivatives of 
the terms relevant in $\delta \beta^{\mu\nu},\delta \beta^{\mu\nu\rho},\delta \beta^{\mu\nu,\rho\sigma}$ 
are less than four in our computations up to the fourth order. 
This means that the fourth derivative terms in $\delta \beta$ can not affect 
$\delta \beta^{\mu\nu},\delta \beta^{\mu\nu\rho},\delta \beta^{\mu\nu,\rho\sigma}$ even in our later computations,
which more or less mixes $\delta \beta,\delta \beta^{\mu\nu},\delta \beta^{\mu\nu\rho},
\delta \beta^{\mu\nu,\rho\sigma}$. 
Therefore, the fourth derivative terms in $\delta \beta$ can be ignored consistently, 
if one is not interested in them: 
our interest 
is up to the second order of derivatives in $\delta \beta$. 

Even with these upper bounds of our interest on the number of derivatives, the computation of \eq{eq:delpf3} is 
very complicated, and we used a Mathematica package ``xTensor" for the 
tensorial computations.
The details of the procedure is explained in Appendix~\ref{app:explicit}.
We have obtained
\[
\begin{split}
\delta \beta^{\mu\nu,\rho\sigma}&= 11n \beta\beta^{\mu\nu,\rho\sigma}+4n \beta \beta^{\mu(\rho,\sigma)\nu}
+4 n \beta^{\mu\nu} \beta^{\rho\sigma}+3 p\, n \beta^{(\mu\nu}\beta^{\rho\sigma)}+{\cal O}(\nabla^2) ,\\
\delta \beta^{\mu\nu\rho}&= -14 n \beta^{(\mu\nu,\rho)\sigma} \beta_{,\sigma}
-4 p n \beta^{(\mu\nu}\beta^{\rho)\sigma}_{,\sigma}+4(1-p) n \beta^{(\mu\nu}_{,\sigma}\beta^{\rho)\sigma}
-2n \beta \beta^{(\mu\nu,\rho)\sigma}_{,\sigma}\\
&\ \ \ +4 (1-p) \beta^{(\mu\nu}\beta^{\rho)\sigma} n_{,\sigma}-8 \beta \beta^{(\mu\nu,\rho)\sigma} n_{,\sigma}+{\cal O}(\nabla^3), \\
\delta \beta^{\mu\nu} &=15 n \beta \beta^{\mu\nu}-2(1+p)n \beta^{\mu\nu}_{,\rho} \beta^{\rho\sigma}_{,\sigma}
+2 (1-p) \beta^{\mu\nu} \beta^{\rho\sigma}_{,\sigma} n_{,\rho}-2 p n \beta^{\mu\rho}_{,\sigma}\beta^{\nu\sigma}_{,\rho} \\
&\ \ \ 
+4 (1-p) \beta^{\rho(\mu} \beta^{\nu)\sigma}_{,\rho}n_{,\sigma}
+4(1-p)\beta^{\rho(\mu}\beta^{\nu)\sigma}_{,\sigma} n_{,\rho}+2 (1-p) \beta^{\rho\sigma} \beta^{\mu\nu}_{,\rho} n_{,\sigma}
-2 p n \beta^{\mu\rho}_{,\rho}\beta^{\nu\sigma}_{,\sigma}\\
&\ \ \ -10 n \beta_{,\rho} \beta^{\mu\nu,\rho\sigma}_{,\sigma}
 -4 \beta \beta^{\mu\nu,\rho\sigma}_{,\rho} n_{,\sigma}-8n\beta_{,\rho} \beta^{\rho(\mu,\nu)\sigma}_{,\sigma}
-8 \beta \beta^{\rho(\mu,\nu)\sigma}_{,\rho} n_{,\sigma}-4 \beta_{,\rho} \beta^{\mu\nu,\rho\sigma}n_{,\sigma}
\\
&\ \ \ -8 \beta^{\rho(\mu,\nu)\sigma}\beta_{,\rho} n_{,\sigma}
-2n \beta_{,\rho\sigma} \beta^{\mu\nu,\rho\sigma}-4n \beta_{,\rho\sigma} \beta^{\mu\rho,\nu\sigma}
+(1-p) n \beta^{\mu\nu} \beta^{\rho\sigma}_{,\rho\sigma}\\
&\ \ \ +4 (1-p) n \beta^{\rho(\mu}\beta^{\nu)\sigma}_{,\rho\sigma}
+(2-p) n \beta^{\rho\sigma} \beta^{\mu\nu}_{,\rho\sigma}+n \beta \beta^{\mu\nu,\rho\sigma}_{,\rho\sigma}
-4n \beta \beta^{\rho(\mu,\nu)\sigma}_{,\rho\sigma}\\
&\ \ \ +(6-p) \beta^{\mu\nu}\beta^{\rho\sigma}n_{,\rho\sigma}
+(4-2p) \beta^{\mu\rho}\beta^{\nu\rho} n_{,\rho\sigma}
+7\beta \beta^{\mu\nu,\rho\sigma}n_{,\rho\sigma}-4 \beta \beta^{\mu\rho,\nu\sigma}n_{,\rho\sigma}\\
&\ \ \ 
+n\left(\frac{4}{3} \beta^{\rho\sigma}\beta^{\delta(\mu}+2 \beta \beta^{\rho\sigma,\delta(\mu}\right)R^{\nu)}{}_{\rho\sigma\delta}
+{\cal O}(\nabla^4),\\
\delta \beta&= 9 n \beta^2 - 4n \beta_{,\mu}\beta^{\mu\nu}_{,\nu}+n \beta^{\mu\nu}\beta_{,\mu\nu}
+n \beta\beta^{\mu\nu}_{,\mu\nu}+5 \beta \beta^{\mu\nu}n_{,\mu\nu}+{\cal O}(\nabla^4),
\end{split}
\label{eq:explicitbetas}
\]
where $p=\frac{4}{3}$ must be taken\footnote{The parameter $p$ becomes a free parameter
in the case that the term $\delta \beta^{\mu\nu\rho\sigma} f^2 f_{,\mu\nu\rho\sigma}$ is also allowed in the expression of 
$\delta P f^3$. 
As explained in Appendix~\ref{app:fourth}, this term can be set to zero by using \eq{eq:ambiguity}
for the unique representation. 
But, if we leave it, $\delta \beta^{\mu\nu\rho\sigma}=(2-3p/2)n \beta^{(\mu\nu}\beta^{\rho\sigma)}$, 
and the others will be given by \eq{eq:explicitbetas} with free $p$.}.
The round brackets in the indices represent symmetrization of the indices contained in the pairs of the brackets.
For example,
$\beta^{\mu(\nu,\rho)\sigma}=\frac{1}{2} \left(  \beta^{\mu\nu,\rho\sigma}+\beta^{\mu\rho,\nu\sigma}\right)$, and $\beta^{(\mu\nu}
\beta^{\rho\sigma)}$ represent the total symmetrization.

As seen in \eq{eq:explicitbetas}, $\delta \beta$'s have complicated expressions with the derivatives of 
both $\beta$'s and $n$. 
The existence of the derivatives of $n$ seems to pose a challenge in comparison with general relativity, 
since the equation of motion 
of the metric tensor field  in the Hamilton-Jacobi formalism of general relativity, 
written down in Section~\ref{sec:conttheory}, contains no derivatives of the lapse function.
This absence comes from the fact that the Hamiltonian of the ADM 
formalism $H_{ADM}$ is expressed with no derivatives of the lapse function, and the Poisson brackets
with the fields do not produce them either, 
where the conjugate momenta to the fields are replaced by some functions of the fields in the Hamilton-Jacobi 
formalism.

The fundamental reason why we encounter the above difference between CTM and general relativity can intuitively be understood
by the fact that, in CTM, a space is an emergent object characterized by the tensor $P$. 
As explained at the beginning of Section~\ref{sec:repP}, there exists intrinsic fuzziness
which disturbs the exactness of a position specified by the coordinate $x$, where 
the ambiguity would be in the order of $\sim\sqrt{\beta^{\mu\nu}/\beta}$ for a dimensional reason.
This ambiguity of positions would also make ambiguous the value of a field, here the lapse function, 
as a function of $x$ by an amount in the order of $\delta n(x)\sim \beta^{\mu\nu}n_{,\mu\nu}/\beta$. 
The real expressions in \eq{eq:explicitbetas} are much more involved, but this gives an 
intuitive understanding of the reason why the spatial derivatives of the lapse function can appear,
irrespective of their absence in general relativity. Therefore, to make relations between CTM and general relativity, 
it would be natural to perform some redefinitions of the lapse function and the fields
by adding some corrections of the spacial derivatives.
In fact, we will do so in later sections.

Another interesting thing to notice in \eq{eq:explicitbetas}
is that there appear terms with the background curvature in $\delta \beta^{\mu\nu}$.
For a static background considered in this section, the background curvature appears just as the coefficients 
of the quadratic terms 
of $\beta$'s, and do not seem to play important roles. 
On the other hand, as we will discuss in later sections, when the background metric
becomes dynamical as a result of the gauge-fixing to a combination of the fields, 
the curvature terms play essential roles for the consistency of the time evolution. 

The result \eq{eq:explicitbetas} shows that there appears a spin-three component $\delta \beta^{\mu\nu\rho}$,
even if we assume $\beta^{\mu\nu\rho}=0$ initially. Therefore, as explained in Section~\ref{sec:gauge},
to maintain the gauge condition $\beta^{\mu\nu\rho}=0$, the spin-three gauge transformation $\delta_3$ 
in \eq{eq:delta3} has to be performed simultaneously.
This is to bring in the spin-three gauge transformation contained in the SO$({\cal N})$ 
rotation part of EOM \eq{eq:ctmeom}.
By setting $\delta \beta^{\mu\nu\rho}+3 \beta v^{\mu\nu\rho}=0$,
we obtain the EOM for the fields as
\[
\begin{split}
&\dot \beta=\delta\beta, \\
&\dot \beta^{\mu\nu}=\delta \beta^{\mu\nu}-\frac{3}{2} \beta\ \nabla_\rho \left(\frac{1}{\beta}\delta\beta^{\mu\nu\rho}\right), \\
&\dot \beta^{\mu\nu,\rho\sigma}=\delta \beta^{\mu\nu,\rho\sigma},
\end{split}
\label{eq:bareeom}
\]
where the last term in the second line comes from the second line of \eq{eq:delta3}, 
the consequence of maintaining the gauge fixing condition $\beta^{\mu\nu\rho}=0$.

A physically important consistency check of the EOM \eq{eq:bareeom}
is to compute the commutation of two successive 
infinitesimal time evolutions. This corresponds to the commutation of the Hamiltonian constraints in CTM,
and, from the first-class nature of the constraint algebra, this should be described by the kinematical 
transformation ${\cal J}_{ab}$. 
In the present context of the continuum limit, the commutation of the time evolutions 
should be expressed by the gauge transformations discussed in the 
preceding section. Since the spin-three transformation $\delta_3$ has already been used for the gauge fixing, 
one would expect that the commutation should be described by the diffeomorphism transformation $\delta_1$.
Note that the lapse function $n(x)$ is a field locally depending on $x$, and the situation is 
the same as the time evolution in terms of the Hamiltonian constraint in general relativity: 
the commutation of Hamiltonian constraint being equal to the diffeomorphism is nothing but the assurance of 
the spacetime covariance of the locally generated time evolution. 
This is directly connected to the central principle in general relativity, 
and it is highly interesting to check this in the present context.

Now, let us explicitly describe the commutation of two successive infinitesimal time evolutions. 
Suppose we start with a configuration,
$\beta,\beta^{\mu\nu},\beta^{\mu\nu,\rho\sigma}$.
After an infinitesimal time $\Delta t$ with lapse $n_1$, the fields evolve to
\[
\beta_1^i=\beta^i+\Delta t \, \dot \beta^i(n_1,\beta,\beta^{\mu\nu},\beta^{\mu\nu,\rho\sigma}),
\label{eq:beta1}
\]
where $\beta^i$ represents $\beta, \beta^{\mu\nu}$, or $\beta^{\mu\nu,\rho\sigma}$. 
Here, we have written explicitly the dependence of $\dot \beta$'s on $n$ and $\beta$'s.
Then, after the second step with lapse $n_2$, 
we obtain
\[
\beta^i_{12}=\beta^i_1+ \Delta t \, \dot \beta^i(n_2,\beta_1,\beta_1^{\mu\nu},\beta_1^{\mu\nu,\rho\sigma}).
\label{eq:beta21}
\]
By inserting \eq{eq:beta1} into \eq{eq:beta21}, expanding in the infinitesimal parameter $\Delta t$, and 
subtracting the case that $n_1$ and $n_2$ are interchanged,  
one obtains
\[
\begin{split}
(\delta_{n_1} \delta_{n_2}-\delta_{n_2}\delta_{n_1})\beta^i&=
\beta^i_{12}-\beta^i_{21}\\
&=(\Delta t)^2 \int d^d x\, \dot \beta^j(x,n_1,\beta,\ldots) 
\frac{\delta}{\delta \beta^j(x)} \dot \beta^i(n_2,\beta,\ldots) -(n_1 \leftrightarrow n_2),
\end{split}
\label{eq:del12m21}
\]
where $j$ is summed over, and we have taken the lowest non-trivial order in $\Delta t$.

We have used ``xTensor" to obtain the following explicit result of \eq{eq:del12m21}: 
\[
(\delta_{n_1} \delta_{n_2}-\delta_{n_2}\delta_{n_1})\beta^i=\delta_1 \beta^i+{\cal O}(\nabla^4),
\label{eq:n1n2v}
\]
where we have dropped the infinitesimal parameter  $\Delta t$, 
$\beta^i=\beta$ or $\beta^{\mu\nu}$, and $\delta_1$ is the diffeomorphism transformation \eq{eq:diffeo} with
\[
v^{\mu}=12 \beta \beta^{\mu\nu}\left( n_1 n_{2,\nu}-n_2 n_{1,\nu} \right).
\label{eq:vval}
\]
The case with $\beta^i=\beta^{\mu\nu,\rho\sigma}$ is not considered, because this requires a higher order computation
than the fourth.
If we make the identification 
\[
\frac{g^{\mu\nu}}{\sqrt{g}}=\beta \beta^{\mu\nu},
\label{eq:geqbeta2}
\]
the commutation algebra \eq{eq:n1n2v} with \eq{eq:vval} agrees with that of the ADM formalism of general relativity
except for a weight factor $1/\sqrt{g}$.
The weight factor is necessary for the consistency with the weights of $\beta$ and $\beta^{\mu\nu}$
shown in \eq{eq:weightbeta}.  
The identification \eq{eq:geqbeta2} was first discussed in \cite{Sasakura:2015pxa} 
with a different argument directly taking the formal continuum limit of the constraint algebra,
and the extra weight factor has been interpreted consistently. 
In Section~\ref{sec:identify}, we will use this relation \eq{eq:geqbeta2} to gauge-fix the background metric,
and the issue of weights will be treated in Section~\ref{sec:delete}.

It is worth mentioning that there exists a scale invariance in the EOM \eq{eq:bareeom} with \eq{eq:explicitbetas}.
The transformation is given by
\[
\begin{split}
t&\rightarrow Lt,\ x^\mu\rightarrow L x^\mu,\\
\beta&\rightarrow \frac{\beta}{L},\ \beta^{\mu\nu}\rightarrow L \beta^{\mu\nu},\  
\beta^{\mu\nu\rho}\rightarrow L^2 \beta^{\mu\nu\rho},\ \beta^{\mu\nu,\rho\sigma}\rightarrow L^3 \beta^{\mu\nu,\rho\sigma},
\end{split}
\label{eq:scaletrans}
\]
where $L$ is a real free parameter.  
The lapse function $n$ and the inverse metric $g^{\mu\nu}$ do not transform. 
The transformation is consistent with the identification \eq{eq:geqbeta2}.
This scale invariance will be respected throughout this paper in the other forms of EOM which will appear in due course.

Lastly, we will present a solution to the EOM for the highest component $\beta^{\mu\nu,\rho\sigma}$. 
Let us assume the following form of a solution,
\[
\beta^{\mu\nu,\rho\sigma}=\frac{a}{\beta}\beta^{\mu\nu}\beta^{\rho\sigma}+\frac{b}{\beta} \beta^{(\mu\nu}\beta^{\rho\sigma)},
\label{eq:ansatzbe4}
\]
where $a,b$ are real numbers. Note that the form is consistent with the scale transformation \eq{eq:scaletrans}.
To check whether this satisfies the EOM,
it is enough to compute the time-derivative of the right-hand side of \eq{eq:ansatzbe4} 
up to non-derivative terms, since we consider $\beta^{\mu\nu,\rho\sigma}$ up to the zeroth order. 
Since, from \eq{eq:bareeom},
\[
\begin{split}
\dot \beta&=9 n \beta^2 +\hbox{derivative terms}, \\
\dot \beta^{\mu\nu}&=15 n \beta \beta^{\mu\nu}+\hbox{derivative terms},
\end{split}
\label{eq:zeroth}
\]
one obtains
\[
\dot \beta^{\mu\nu,\rho\sigma}=21 n \left(a \beta^{\mu\nu}\beta^{\rho\sigma}+b \beta^{(\mu\nu}\beta^{\rho\sigma)} \right)
+\hbox{derivative terms}
\]
from the assumption \eq{eq:ansatzbe4}. 
On the other hand, by inserting \eq{eq:ansatzbe4} into the EOM 
\eq{eq:bareeom}, one obtains
\[
\dot \beta^{\mu\nu,\rho\sigma}=(9a+4) n \beta^{\mu\nu}\beta^{\rho\sigma}+(6a+15b+4)n\beta^{(\mu\nu}\beta^{\rho\sigma)}
+\hbox{derivative terms}.
\]
By equating the two expressions for $\dot \beta^{\mu\nu,\rho\sigma}$, one obtains
\[
a=\frac{1}{3},\ b=1.
\label{eq:valab}
\]
The existence of the consistent solution implies that one can 
ignore the field $\beta^{\mu\nu,\rho\sigma}$ assuming that 
it is given by \eq{eq:ansatzbe4} with \eq{eq:valab}. 
This truncation for simplicity will be assumed in the further analysis in later sections.

\section{Gauge symmetry of the background metric}
\label{sec:backmetric}
\begin{figure}
\begin{center}
\includegraphics[scale=1]{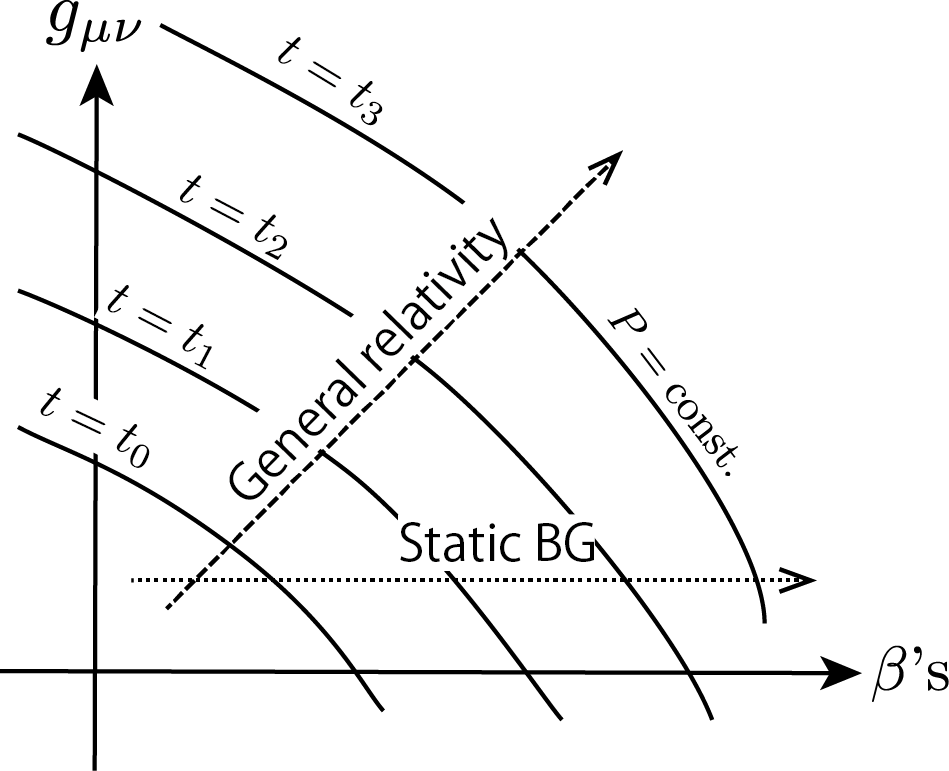}
\caption{A schematic illustration of the time evolution in CTM. The horizontal and vertical axes 
represent the configurations of the fields and the background metric, respectively. 
The solid curves represent the submanifolds of constant $P$.
A time evolution is a transition from a constant $P$ 
submanifold to another in the configuration space.
The dotted arrow represents a time evolution in the gauge of a static background metric, while the 
dashed arrow represents an evolution which describes time evolution 
in general relativity by the gauge choice \eq{eq:geqbeta2}.}
\label{fig:gauge}
\end{center}
\end{figure}

In the former sections, we considered a static background metric, and this is certainly a consistent treatment. 
However, there exist two distinct rank-two symmetric tensors, $g^{\mu\nu}$ and $\beta^{\mu\nu}$, and 
this would be physically awkward from the view point of general relativity, which has a unique symmetric rank-two
tensor called the metric.
In fact, as will be explained below, the background metric can be chosen arbitrarily without changing the physical contents
of CTM: there exists a gauge symmetry which allows one to freely change the background metric
with compensation by the fields.
In other words, as illustrated in Figure~\ref{fig:gauge},
a constant surface of $P$ forms a submanifold in the configuration space of $g_{\mu\nu}$ and $\beta$'s,
and it is extending in the directions that allow arbitrary infinitesimal changes of the background metric.  
Since the motion of $P$ is determined by $P$ itself as in \eq{eq:ctmeom} (up to the kinematical gauge symmetry), 
the motion is actually a time-dependent transition from a constant $P$ submanifold to another.  
Such transitions can be described by various manners of one's own choice, as illustrated 
for two examples in Figure~\ref{fig:gauge}.
Taking a representative point on each constant $P$ submanifold determines a trajectory of time evolution
in the configuration space of $g_{\mu\nu}$ and $\beta$'s.  This 
is a gauge choice, and, in the former section, we take the gauge that the background metric is static, and the motion is solely 
described by $\beta$'s.
This is illustrated as the dotted arrow in the figure. On the other hand, we may take another choice that $g_{\mu\nu}$
and $\beta$'s are correlated. This is what we will take for the comparison with general relativity, in which the actual gauge fixing
condition will be taken as \eq{eq:geqbeta2}. This is illustrated as a dashed arrow in the figure. 
Note that the two descriptions are physically equivalent: they are connected by a transformation 
of $g_{\mu\nu}$ and $\beta$'s along a constant $P$ submanifold, while $g_{\mu\nu}$ and $\beta$'s 
take different values.

Let us describe the submanifold of constant $P$ by considering the infinitesimal changes of $g_{\mu\nu}$ and 
$\beta$'s which keep $P$.  This condition is given by $\delta (Pf^3)=0$ with 
the test function unchanged $\delta f=0$, while $g_{\mu\nu}$ and $\beta$'s  are allowed to be changed. 
By taking the infinitesimal of \eq{eq:pf3}, it is straightforward to derive
\[
\begin{split}
\delta (Pf^3)=\int d^d x & \left[
\left(\delta \beta -\frac{1}{2} \beta^{\mu\nu} \delta\Gamma_{\mu,\nu}
+\frac{1}{3} \left(  \beta^{\mu\nu}\delta\tilde\Gamma_{\mu\nu}^\rho\right)_{,\rho} \right)f^3 \right.\\
&
+\left(
\delta \beta^{\mu\nu} -\beta^{\mu\nu,\rho\sigma} \delta\Gamma_{\rho,\sigma}
+\left(\beta^{\mu\nu,\rho\sigma} \delta \tilde \Gamma_{\rho\sigma}^\delta \right)_{,\delta}
\right)f^2 f_{,\mu\nu} 
\\
&+\left.
\left(\delta \beta^{\delta \rho\sigma}+\beta^{\mu\nu,\rho\sigma}\delta \tilde \Gamma_{\mu\nu}^\delta 
\right)f^2 f_{,\delta\rho\sigma}
+\delta \beta^{\mu\nu,\rho\sigma}f f_{,\mu\nu} f_{,\rho\sigma}
\right]+{\cal O}(\nabla^5),
\end{split}
\label{eq:delpf3met}
\]
where 
\[
\tilde \Gamma_{\mu\nu}^\rho:= \Gamma_{\mu\nu}^\rho+ \delta^{\rho}_{(\mu}\Gamma_{\nu)},
\] 
and 
\[
\delta \Gamma_{\mu\nu}^\rho=\frac{1}{2} g^{\rho\sigma}\left(
\nabla_\mu \delta g_{\nu\sigma}+\nabla_\nu \delta g_{\mu\sigma}-\nabla_\sigma \delta g_{\mu\nu}\right).
\]
Here, we have assumed the gauge condition $\beta^{\mu\nu\rho}=0$ as an initial input.
To derive the result, we have considered the change of the covariant derivatives under the change of $g_{\mu\nu}$, 
namely,
\[
\delta f_{,\mu\nu}=-\delta\Gamma_{\mu\nu}^\rho f_{,\rho}-\frac{1}{2} \delta \Gamma_\mu f_{,\nu}
+\nabla_{\mu} \left(- \frac{1}{2}\delta\Gamma_\nu f \right),
\]
and have performed some partial integrations to obtain \eq{eq:delpf3met}.
To further transform it to the form \eq{eq:pf3}, we have to symmetrize the third derivative of $f$ 
by using the following equation with the Riemann tensor: 
\[
\begin{split}
\beta^{\mu\nu,\rho\sigma}\delta \tilde \Gamma_{\mu\nu}^\delta f^2 f_{,(\delta\rho\sigma)}
=&\frac{1}{3}\beta^{\mu\nu,\rho\sigma}\delta \tilde \Gamma_{\mu\nu}^\delta f^2
\left(f_{,\delta\rho\sigma}+f_{,\rho\delta\sigma}+f_{,\rho\sigma\delta}\right)\\
=&\frac{1}{3}\beta^{\mu\nu,\rho\sigma}\delta \tilde \Gamma_{\mu\nu}^\delta f^2 
\left(3f_{,\delta\rho\sigma}+2 R_{\rho\delta\sigma}{}^\kappa f_{,\kappa}\right).
\end{split}
\]
The last term in the last line can be transformed to the non-derivative terms of $f$ by a partial integration, because 
$f^2 f_{,\kappa}=\frac{1}{3} (f^3)_{,\kappa}$. Then, the condition $\delta (Pf^3)=0$ implies
\[
\begin{split}
&\delta \beta=\frac{1}{2} \beta^{\mu\nu} \delta\Gamma_{\mu,\nu}
-\frac{1}{3} \left(  \beta^{\mu\nu}\delta\tilde\Gamma_{\mu\nu}^\rho\right)_{,\rho} 
-\frac{2}{9}\left( \beta^{\mu\nu,\rho\sigma}\delta \tilde \Gamma_{\mu\nu}^\delta R_{\rho\delta\sigma}{}^\kappa\right)_\kappa, \\
&\delta \beta^{\mu\nu}= \beta^{\mu\nu,\rho\sigma} \delta\Gamma_{\rho,\sigma}
-\left(\beta^{\mu\nu,\rho\sigma} \delta \tilde \Gamma_{\rho\sigma}^\delta \right)_{,\delta} ,\\
&\delta \beta^{\mu\nu\rho}=-\beta^{\sigma\delta,(\mu\nu}\delta \tilde \Gamma_{\sigma\delta}^{\rho)} ,\\
&\delta \beta^{\mu\nu,\rho\sigma}=0.
\end{split}
\label{eq:deformbeta}
\]
We have shown that an arbitrary infinitesimal deformation of the background metric can be absorbed by
the infinitesimal change of $\beta$'s shown in \eq{eq:deformbeta}.
The last term in the first line is actually irrelevant, because it is higher order than our range of interest.
Note that there appear spin-3 components, which must be absorbed in the way discussed in Section~\ref{sec:gauge}
to maintain the gauge condition $\beta^{\mu\nu\rho}=0$.

\section{Identifying the background geometry with the fields} 
\label{sec:identify}
The background geometry introduced in the preceding sections is arbitrary. 
In fact, as discussed in Section~\ref{sec:backmetric},
an arbitrary change of the background geometry can be absorbed into the change of the fields $\beta$'s
without changing $P$. This means that there exists a gauge symmetry which changes the background geometry
without changing the dynamical contents of the system.
The most reasonable choice of the background geometry is \eq{eq:geqbeta2}, which 
determines the background geometry in terms of $\beta$ and $\beta^{\mu\nu}$, and 
makes it a dynamical entity. 

If we impose the identification \eq{eq:geqbeta2},
the diffeomorphism transformation \eq{eq:diffeo} derived previously for a static background will also be changed.
This is because we have to take into account the simultaneous transformation of $g^{\mu\nu}$ keeping the 
relation \eq{eq:geqbeta2}.
It is easy to see that the corrections are given by the minus of \eq{eq:deformbeta}. 
Therefore, since \eq{eq:diffeo} is in the first order of derivatives,
the corrections are higher than the second order of derivatives.
This is out of our range of interest, and the diffeomorphism transformation 
remains in the form \eq{eq:diffeo}. 
This is consistent with the naive expectation that
$\beta$ and $\beta^{\mu\nu}$ should still behave as a scalar and a two-tensor with the weight of negative half-density,
even after the identification of the background metric with the fields.

It is important to see whether the transformation \eq{eq:diffeo} and the identification \eq{eq:geqbeta2} 
reproduce the standard diffeomorphism transformation of $g^{\mu\nu}$. Let us define
\[
\tilde g^{\mu\nu}\equiv \beta\beta^{\mu\nu}=\frac{g^{\mu\nu}}{\sqrt{g}},
\label{eq:deftildeg}
\]
where we wrote \eq{eq:geqbeta2} as well. From \eq{eq:diffeo}, one obtains
\[
\delta_1 \tilde g^{\mu\nu}&=(\delta_1 \beta)\beta^{\mu\nu}+\beta (\delta_1 \beta^{\mu\nu})\CR
&=-v^\rho \tilde g^{\mu\nu}_{,\rho}+v^\rho_{,\rho} \tilde g^{\mu\nu}+v^\mu_{,\rho}\tilde g^{\rho\nu}+
v^\nu_{,\rho}\tilde g^{\mu\rho}+{\cal O}(\nabla^3).
\label{eq:deltagtilde}
\]
Then, by using the second relation in \eq{eq:deftildeg}, one obtains
\[
\begin{split}
\delta_1 g^{\mu\nu}&=\sqrt{g}\left( \delta_1 \tilde g^{\mu\nu}-\frac{g^{\mu\nu}}{d+2} g_{\rho\sigma} \delta_1 \tilde g^{\rho\sigma}
\right) \\
&=\nabla^\mu v^\nu+\nabla^\nu v^\mu+{\cal O}(\nabla^3 ),
\end{split}
\]
where we have put \eq{eq:deltagtilde}.
This indeed agrees with the transformation of the metric under the diffeomorphism in general relativity.

One consequence of the identification \eq{eq:geqbeta2} is that the expression of $\delta \beta$'s 
in \eq{eq:explicitbetas} is considerably simplified. This comes from $\nabla_\mu (\beta \beta^{\nu\rho})=0$, 
which is because the covariant derivative satisfies $\nabla_\mu g^{\nu\rho}=0$. 
By substituting \eq{eq:explicitbetas} with \eq{eq:geqbeta2}, \eq{eq:ansatzbe4} and \eq{eq:valab}, 
we obtain
\[
\begin{split}
\tilde \delta \beta&=9n \beta^2+\frac{6n}{\beta^2} \tilde g^{\mu\nu} \beta_{,\mu}\beta_{,\nu}+5 \tilde g^{\mu\nu} n_{,\mu\nu}
+{\cal O}(\nabla^4),\\
\tilde \delta \beta^{\mu\nu}&= 15 n \tilde g^{\mu\nu} -\frac{20n}{\beta^4} \tilde g^{\mu\rho}\tilde g^{\nu\sigma} \beta_{,\rho}
\beta_{,\sigma}-\frac{8}{\beta^3} \tilde g ^{\rho(\mu} \tilde g^{\nu)\sigma}\beta_{,\rho} n_{,\sigma}
+\frac{10 n}{\beta^3} \tilde g^{\rho\mu}\tilde g^{\nu\sigma} \beta_{,\rho\sigma}
+\frac{14}{\beta^2} \tilde g^{\mu\rho}\tilde g^{\nu\sigma} n_{,\rho\sigma}\\
&\ \ \ +\tilde g^{\mu\nu}\left(\frac{8n}{\beta^4} \tilde g^{\rho\sigma} \beta_{,\rho}\beta_{,\sigma} 
-\frac{4}{\beta^3} \tilde g ^{\rho\sigma} \beta_{,\rho} n_{,\sigma}
+\frac{n}{\beta^3} \tilde g^{\rho\sigma} \beta_{,\rho\sigma}
+\frac{14}{\beta^2} \tilde g^{\rho\sigma} n_{,\rho\sigma}
\right) -\frac{2n}{\beta^2} \tilde g^{\mu\rho}\tilde g^{\nu\sigma} R_{\rho\sigma}+{\cal O}(\nabla^4).
\end{split}
\label{eq:tildebeta}
\]
Here, note that $\tilde \delta \beta^{\mu\nu\rho}$ and $\tilde \delta \beta^{\mu\nu,\rho\sigma}$ are not considered anymore:
$\tilde \delta \beta^{\mu\nu\rho}$ has been gauged away to be included in $\tilde \delta \beta^{\mu\nu}$ by
the spin-three gauge transformation to keep the gauge condition $\beta^{\mu\nu\rho}=0$,
and $\beta^{\mu\nu,\rho\sigma}$ is assumed to be the solution \eq{eq:ansatzbe4} with \eq{eq:valab}.

As can be seen in \eq{eq:tildebeta}, while the right-hand sides of the equation of motion (EOM) have considerably been simplified
in comparison with \eq{eq:explicitbetas},
the left-hand side, $\frac{d}{dt} (Pf^3)$, 
must be modified with some additional terms which come from the evolution of the background metric to 
keep the relation \eq{eq:geqbeta2}:
the left-hand side can not simply be expressed by the time-derivatives of the fields 
$\dot \beta,\dot \beta^{\mu\nu}$, but must also contain some additional terms coming from the time-derivative of $g_{\mu\nu}$
contained in the covariant derivatives in \eq{eq:pf3}.
The derivation of the explicit expression of the left-hand side is basically the same as that of \eq{eq:deformbeta}
through \eq{eq:delpf3met},
and the additional terms are just the minus of the right-hand sides of \eq{eq:deformbeta} with the replacement 
$\delta \Gamma\rightarrow \dot \Gamma$. In addition, to keep the gauge condition $\beta^{\mu\nu\rho}=0$, we
have to perform the spin-three gauge transformation to transfer 
$\delta \beta^{\mu\nu\rho}$ in \eq{eq:deformbeta} to  
$\dot \beta^{\mu\nu}$.  Then, we obtain the EOM as
\[
\begin{split}
&\dot \beta-\frac{1}{2} \beta^{\mu\nu} \dot \Gamma_{\mu,\nu}+\frac{1}{3} \nabla_{\sigma}
\left(\beta^{\mu\nu} \dot{\tilde\Gamma}_{\mu\nu}^\sigma \right)=\tilde \delta \beta,\\
&\dot \beta^{\mu\nu}-\beta^{\mu\nu,\rho\sigma}\dot \Gamma_{\rho,\sigma}+
\nabla_\delta \left(\beta^{\mu\nu,\rho\sigma}\,\dot{\tilde \Gamma}_{\rho\sigma}^\delta \right)
-\frac{3}{2}\beta \nabla_\rho \left(\frac{1}{\beta} \beta^{\sigma\delta,(\mu\nu} \dot{\tilde \Gamma}_{\sigma\delta}^{\rho)}\right)
=\tilde \delta \beta^{\mu\nu},
\end{split}
\label{eq:eommiddle}
\]
where \eq{eq:geqbeta2}, \eq{eq:ansatzbe4} and \eq{eq:valab} are supposed, and the last term on the left-hand side of the 
last line comes from the spin-three transformation.
It would be worth to remind that the time derivative of the Christoffel symbol can be written covariantly as 
\[
\dot \Gamma_{\mu\nu}^\rho=\frac{1}{2} g^{\rho\sigma}\left( \nabla_\mu \dot g_{\nu\sigma}+\nabla_\nu \dot g_{\mu\sigma}-
\nabla_\sigma \dot g_{\mu\nu}\right),
\label{eq:covgam}
\]
and therefore \eq{eq:eommiddle} is a covariant expression.

Let us simplify \eq{eq:eommiddle} further.
In the zeroth order of derivatives, the equation of motion (EOM) derived from \eq{eq:eommiddle} is still given by \eq{eq:zeroth}, 
since all the corrections in \eq{eq:eommiddle} are in the second order. Therefore, by using \eq{eq:geqbeta2}, 
the EOM of $g^{\mu\nu}$ in the zeroth order is given by
\[
\dot g^{\mu\nu}=\frac{48 n \beta}{d+2}  g^{\mu\nu}+{\cal O}(\nabla^2).
\label{eq:dotg}
\]
Here, the dimensional dependence appears due to
the determinant in \eq{eq:geqbeta2}, while the EOM so far has been independent of it. Then, by
putting \eq{eq:dotg} into \eq{eq:covgam}, one obtains
\[
\dot\Gamma_{\mu\nu}^\rho=-\frac{48}{d+2} 
\left(\delta^\rho_{(\mu}\nabla_{\nu)}(n \beta)-\frac{1}{2}g_{\mu\nu} \nabla^\rho(n\beta)\right)+{\cal O}(\nabla^3).
\label{eq:dotgamma}
\]
The overall minus sign is from the fact $\dot g_{\mu\nu}=-g_{\mu\rho}g_{\nu\sigma} \dot g^{\rho\sigma}$.
This order of $\dot \Gamma$ is enough for our second order computation of 
the correction terms on the left-hand side of \eq{eq:eommiddle}. 
By putting \eq{eq:dotgamma} into \eq{eq:eommiddle}, we finally obtain
\[
\begin{split}
&\dot \beta
=9 n \beta^2+\frac{\tilde g^{\mu\nu}}{d+2}\left(\frac{2(3d-2)n}{\beta^2} \beta_{,\mu} \beta_{,\nu}
-\frac{8(3d-2)}{\beta}  \beta_{,\mu}n_{,\nu}
-\frac{4(3d-4)n}{\beta}  \beta_{,\mu\nu}
-(7d-26) n_{,\mu\nu} \right)  \\
&\ \ \ \ \ \  +{\cal O}(\nabla^4), \\
&\dot \beta^{\mu\nu}=15 n \tilde g^{\mu\nu} -\frac{2n \tilde g^{\mu\rho} \tilde g^{\nu\sigma}}{\beta^2}R_{\rho\sigma} \\
&\ \ \ \ \ \  +\frac{\tilde g^{\rho(\mu}\tilde g^{\nu)\sigma}}{(d+2)\beta^2}\left( -\frac{4(d-14)n}{\beta^2}\beta_{,\rho}\beta_{,\sigma}
-\frac{24(d-2)}{\beta} \beta_{,\rho}n_{,\sigma}  
-\frac{2(3d-2)n}{\beta} \beta_{,\rho\sigma}
-2(d-6) n_{,\rho\sigma}
 \right)
\\
&\ \ \ \ \ \  +\frac{\tilde g^{\mu\nu}\tilde g^{\rho\sigma}}{(d+2)\beta^2}\left( \frac{8(5d+8)n}{\beta^2} \beta_{,\rho}\beta_{,\sigma} 
-\frac{4(5d-6)}{\beta}\beta_{,\rho}n_{,\sigma}
-\frac{(23d+6)n}{\beta} \beta_{,\rho\sigma}
-10(d-2) n_{,\rho\sigma} 
\right) \\
&\ \ \ \ \ \ +{\cal O}(\nabla^4),
\end{split}
\label{eq:eombetawithg}
\]
where \eq{eq:geqbeta2} is supposed. This is the version of EOM with a dynamical background metric
determined by \eq{eq:geqbeta2}.

A physically meaningful consistency check of EOM \eq{eq:eombetawithg} is
given by computing the commutation of two successive infinitesimal time evolutions, as 
the algebraic structure \eq{eq:n1n2v} with \eq{eq:vval} has been obtained for the static background case. 
The existence of the gauge symmetry discussed in Section \ref{sec:backmetric}, which 
allows us to freely change the background metric, 
assures the covariance of the time evolution for the evolving background case, too. 
Therefore, we should obtain the same algebraic structure as the static background case. 
However, 
the actual computation for the consistency check is much more complicated and non-trivial than the fixed background case.  
In the second step of the successive infinitesimal time evolutions, one has to compute the time derivative of 
the right-hand side of 
\eq{eq:eombetawithg}.\footnote{\eq{eq:del12m21} corresponds to acting the time-derivative on $\dot \beta$'s.}
In the computation, the main difference from the static background case 
is that we have to take into account the time derivative of the metric as well, which
affects not only the metric itself but also the covariant derivatives and the curvature tensor.
Therefore, 
while the number of terms in \eq{eq:eombetawithg} has substantially been reduced from \eq{eq:explicitbetas}
by the identification \eq{eq:geqbeta2}, there appear a number of new terms in the 
second step, which someway set back the reduction.
One can compute these extra contributions in a similar manner as was done in Section~\ref{sec:backmetric}.
For instance, as for $\beta$, 
\[
\begin{split}
\frac{d}{dt} \beta_{,\mu}&={\dot \beta}_{,\mu}+\frac{1}{2} \dot \Gamma_\mu \beta,\\
\frac{d}{dt} \beta_{,\mu\nu}&={\dot \beta}_{,\mu\nu}-\dot \Gamma_{\mu\nu}^\rho \beta_{,\rho}+\frac{1}{2} 
\dot \Gamma_\mu \beta_{,\nu}+\frac{1}{2} \nabla_\mu(\dot \Gamma_\nu \beta),
\end{split}
\label{eq:dotsecond}
\]
where the terms with $\dot \Gamma_\mu$ are due to the weight of $\beta$'s in \eq{eq:weightbeta}. Here, 
$\dot \Gamma_{\mu\nu}^\rho$ is explicitly given by \eq{eq:dotgamma}. 
As for the curvature tensor, since the curvature is in the second order by itself, it is enough to consider the non-derivative part 
\eq{eq:dotg} of $\dot g_{\mu\nu}$, and we obtain\footnote{The computation is simplified by noticing that
the non-derivative part of \eq{eq:dotg} is just a conformal transformation.} 
\[
\dot R_{\mu\nu}=\frac{24}{d+2}\left((d-2)\nabla_\mu \nabla_\nu (n\beta)+g_{\mu\nu} \nabla^2(n\beta)\right)+{\cal O}(\nabla^4).
\label{eq:dotR}
\]
By using these expressions, one can compute the commutation of infinitesimal time evolutions, and obtain 
\[
\begin{split}
&(\delta_{n_1}\delta_{n_2}-\delta_{n_2}\delta_{n_1})\beta
=-\tilde g^{\mu\nu}v_\mu \beta_{,\nu}+\frac{1}{2} \tilde g^{\mu\nu}v_{\mu,\nu}\beta +{\cal O}(\nabla^4), \\
&(\delta_{n_1}\delta_{n_2}-\delta_{n_2}\delta_{n_1})\beta^{\mu\nu}
=\frac{1}{\beta^2}\left(2\tilde g^{\rho(\mu}\tilde g^{\nu)\sigma}v_{\rho,\sigma}
 \beta+\tilde g^{\mu\nu}\tilde g^{\rho\sigma}\left(\frac{1}{2}  v_{\rho,\sigma} \beta+v_\rho \beta_{,\sigma}\right)\right)
 +{\cal O}(\nabla^4),
 \label{eq:hhbeta}
\end{split}
\]
where 
\[
v_\mu=12 (n_1 n_{2,\mu}-n_2 n_{1,\mu}).
\]
One can easily check that the right-hand sides are the same as \eq{eq:n1n2v} with \eq{eq:vval}, when \eq{eq:geqbeta2} 
is taken into account. 
Thus, the right-hand sides of (\ref{eq:hhbeta}) represent the diffeomorphism transformations, and 
the consistency of the time evolution in the case of the evolving background with \eq{eq:geqbeta2} 
has also been established.

\section{Deletion of the weights}
\label{sec:delete}
So far, the field $\beta$ and the lapse function $n$ have the weights of negative and positive half-densities, respectively.
While these are the natural weights in the framework of CTM, 
scalars with such weights are not standard in general relativity.
Therefore, we want to transform them into simple scalars with no weights. 
At first glance, this seems to be a trivial task by doing the replacement,
$\beta\rightarrow g^{-\frac14} \beta$ and
$n\rightarrow g^\frac{1}{4} n$, in the equation of motion (EOM) \eq{eq:eombetawithg}.
However, while the former is obvious, there is a subtle issue in the latter replacement. 

When we have shown the algebraic relation between the commutation of two infinitesimal time evolutions and 
the diffeomorphism in the preceding sections,
it is implicitly assumed that $n_2$ does not change after the first infinitesimal time evolution with $n_1$, and vice versa.
Namely, the algebraic relation has been shown in the situation that the lapse functions with the weight
of half-density do not change after the infinitesimal time evolutions.
On the other hand, if we do the replacement $n\rightarrow g^\frac{1}{4} n$, and assume that the new lapse functions 
with no weights do not change after a first infinitesimal time evolution, the situation becomes in fact different by the 
evolution of the weight $g^\frac{1}{4}$ from the original one. This means that the commutation of two infinitesimal time evolutions 
is a sum of a diffeomorphism and an infinitesimal time evolution
with the following lapse function:
\[
n_{12}=-\frac{1}{4} g_{\mu\nu}\dot g^{\mu\nu}(n_1)n_2+\frac{1}{4} g_{\mu\nu}\dot g^{\mu\nu}(n_2)n_1.
\label{eq:additional}
\]
Here, we have explicitly written the lapse function dependence of $\dot g^{\mu\nu}$, while it depends also on $\beta$ 
and $g^{\mu\nu}$.
Of course, the appearance of an additional time evolution
is not a breakdown of the framework, because the algebraic closure of the diffeomorphism and 
the infinitesimal time evolution anyway holds. But, this deformed algebraic structure is inconvenient, 
if we want to compare CTM with the ADM formalism of general relativity.

To fix this issue, let us consider the following reparameterization of the lapse function,
\[
n\rightarrow \tilde n =n+h(\beta,g^{\mu\nu},n),
\label{eq:repn}
\] 
where $h$ is a scalar function linear in $n$, 
and is assumed to be in the order of second derivatives.\footnote{A direct way to compensate, 
such as $n\rightarrow g^{-\frac{1}{4}} n$, cannot be taken, because $n$ is supposed to be a scalar with no weights, and its
weight should not be changed.}
The reason for $h$ to be taken in the second order is that we want to keep the result in the main order,
namely, the part expressed by the diffeomorphism. 
Then, the condition to compensate \eq{eq:additional} is given by
\[
\begin{split}
&\int dx \left[ \dot\beta(x,n_1) \frac{\delta}{\delta \beta(x)}
+\dot g^{\mu\nu}(x,n_1)\frac{\delta}{\delta g^{\mu\nu}(x)}\right]h(\beta,g^{\mu\nu},n_2)
-\frac{1}{4} g_{\mu\nu} \dot g^{\mu\nu}(n_1)n_2 -(n_1 \leftrightarrow n_2)\\
&\hspace{13cm}={\cal O}(\nabla^4).
\end{split}
\label{eq:hcond}
\]

Before discussing the solution for $h$ to \eq{eq:hcond}, 
let us first discuss the explicit expressions of the EOM
in the case with no weights.
So, let us leave aside the replacement $n\rightarrow \tilde n$ for the moment.
After the rescaling by the weight factors, i.e., $\beta\rightarrow g^{-\frac14} \beta$ and
$n\rightarrow g^\frac{1}{4} n$, the EOM has the form, 
\[
\begin{split}
g^\frac{1}{4} \frac{d}{dt} \left(g^{-\frac{1}{4}} \beta\right)&=K(\beta,g^{\mu\nu},n), \\
g^\frac{1}{4}\frac{d}{dt} \left( g^{-\frac{1}{4}}\beta^{\mu\nu}\right) &=K^{\mu\nu}(\beta,g^{\mu\nu},n),
\end{split}
\label{eq:gK}
\]
where $K$ and $K^{\mu\nu}$ are given by the right-hand sides of \eq{eq:eombetawithg} with
the formal replacement $\tilde g^{\mu\nu} \rightarrow g^{\mu\nu}$.
The left-hand sides of \eq{eq:gK} can be written in the way,
\[
\left(
\begin{array}{cc}
1 & \frac{1}{4} \beta g_{\rho\sigma} \\
-\frac{1}{\beta^2}g^{\mu\nu} & \frac{1}{\beta} I_{\rho\sigma}^{\mu\nu} +\frac1{4\beta} g^{\mu\nu}g_{\rho\sigma} 
\end{array}
\right)
\left(
\begin{array}{c}
\dot \beta \\
\dot g^{\rho\sigma}
\end{array}
\right),
\label{eq:dotbdotg}
\]
where $I^{\mu\nu}_{\rho\sigma}= \delta^{(\mu}_\rho \delta^{\nu)}_\sigma$, and 
\eq{eq:geqbeta2} has been used.
It is easy to find the inverse of the matrix in \eq{eq:dotbdotg}, and we obtain
\[
\left(
\begin{array}{c}
\dot \beta \\
\dot g^{\mu\nu}
\end{array}
\right)=
\left(
\begin{array}{cc}
c_1 & c_2 \beta^2 g_{\rho\sigma} \\
\frac{c_3}{\beta}g^{\mu\nu} & \beta I_{\rho\sigma}^{\mu\nu} +c_4 \beta g^{\mu\nu}g_{\rho\sigma} 
\end{array}
\right)
\left(
\begin{array}{c}
K(\beta,g^{\mu\nu}, n) \\
K^{\rho\sigma}(\beta,g^{\mu\nu},n)
\end{array}
\right),
\label{eq:eomscalar}
\]
where
\[
c_1=\frac{d+4}{2(d+2)},\ c_2=-\frac{1}{2(d+2)},\ c_3=\frac{2}{d+2},\ c_4=-\frac{1}{d+2}.
\] 

Now let us discuss the replacement $n\rightarrow \tilde n$. 
To solve the condition \eq{eq:hcond} for $h$, let us assume the following form,
\[
h(\beta,g^{\mu\nu},n)=\frac{g^{\mu\nu}}{\beta^2}\left(
z_1 \frac{n \beta_{,\mu}\beta_{,\nu}}{\beta^2}+z_2   \frac{\beta_{,\mu}n_{,\nu}}{\beta}
+z_3 \frac{n\beta_{,\mu\nu}}{\beta}+z_4 n_{,\mu\nu}
\right),
\label{eq:hass}
\]
where $z_i$ are parameters. This form is chosen so that the reparameterization \eq{eq:repn} preserves the
original form of the EOM.
By substituting $\dot g^{\mu\nu}$ in \eq{eq:hcond} with \eq{eq:eomscalar}, 
we find that \eq{eq:hcond} can be solved by
\[
\begin{split}
&(d-6) z_3 + 2 (2 + d) z_4 = \frac{-12 - 44 d + 17 d^2}{6(d+2)},\\
&2 ( d-6) z_1 + (10 + d) z_2 + 
    4 (3 d - 10) z_3 + 8(2 -d) z_4 =\frac{2 (-12 - 4 d + 11 d^2)}{3 (d + 2)}.
 \end{split}
 \label{eq:condz}
 \]
The solutions form a two-parameter family, and any of them can be used for the purpose.

The final form of the EOM with no weights of the field and the lapse function 
is obtained by 
doing the replacement $n\rightarrow \tilde n$ in \eq{eq:eomscalar}.
Because our concern is up to the second order, the replacement 
is effective only in the zeroth order terms in \eq{eq:eomscalar}. 
By explicitly computing \eq{eq:eomscalar}, we obtain
\[
\begin{split}
\dot \beta&=-\frac{3(d-6)}{d+2} \beta^2 \left( n+h(\beta,g^{\mu\nu},n)\right)
+\frac{1}{d+2} nR-\frac{17d^2+20d+36}{(d+2)^2 \beta^2} n \beta_{,\mu}\beta^{,\mu}\\
&\ \ \ \ -\frac{2(d^2+20d-4)}{(d+2)^2 \beta} \beta_{,\mu} n^{,\mu} 
+\frac{11d^2-20 d+60}{2(d+2)^2 \beta}n \beta_{,\mu}^{,\mu}+\frac{3 d^2-20 d+92}{2(d+2)^2}n_{,\mu}^{,\mu}
+{\cal O}(\nabla^4), \\
\dot g^{\mu\nu}&=\frac{48 \beta}{d+2} g^{\mu\nu} \left(n+h(\beta,g^{\rho\sigma},n)\right)-\frac{2}{\beta}nR^{\mu\nu}
+ \frac{2 }{(d+2)\beta}nR g^{\mu\nu}\\
&\ \ \  
-\frac{4(d-14)}{(d+2) \beta^3} n \beta^{,\mu} \beta^{,\nu}-\frac{24(d-2)}{(d+2)\beta^2} n^{,(\mu}\beta^{,\nu)}
-\frac{2(3d-2)}{(d+2)\beta^2} n \beta^{,\mu\nu}
-\frac{2(d-6)}{(d+2)\beta} n^{,\mu\nu}\\
&\ \ \ 
+g^{\mu\nu}\left(\frac{32(3d+2)}{(d+2)^2\beta^3}n \beta_{,\rho}\beta^{,\rho}
-\frac{32(2d-1)}{(d+2)^2\beta^2}n_{,\rho}\beta^{,\rho} -\frac{16(4d-1)}{(d+2)^2\beta^2}n\beta_{,\rho}^{,\rho}
-\frac{16(2d-5)}{(d+2)^2 \beta} n_{,\rho}^{,\rho} \right) \\
& \ \ \ +{\cal O}(\nabla^4).
\end{split}
\label{eq:eomfinal}
\]

For a consistency check of this result, one can compute the commutation of two infinitesimal time evolutions,
as done before.
The basic strategy is the same. In the second step of the infinitesimal time evolution, one has to take the 
time derivative of the right-hand sides of \eq{eq:eomfinal}. Not only the metric itself, but we also take into account 
the time derivative of the second covariant derivatives\footnote{The difference from the previous case \eq{eq:dotsecond}
is the absence of weights, namely, $\dot \Gamma_\mu$ is absent.
Because of this, the first covariant derivatives have no time-dependencies.} and the curvature.
Since our concern is up to the second order, the time-derivative of the Christoffel symbol and the curvature can be evaluated by the zeroth order of $\dot g^{\mu\nu}$, as given in \eq{eq:dotgamma} and \eq{eq:dotR}, respectively.
Then, we obtain  
\[
\begin{split}
&(\delta_{n_1}\delta_{n_2} -\delta_{n_2}\delta_{n_1})\beta=-v^\mu \beta_{,\mu}+{\cal O}(\nabla^4), \\
&(\delta_{n_1}\delta_{n_2} -\delta_{n_2}\delta_{n_1})g^{\mu\nu}=2v^{(\mu,\nu)}+{\cal O}(\nabla^4),
\end{split}
\] 
where 
\[
v^\mu=12\left( n_1 n_2^{,\mu}-n_2 n_1^{,\mu}\right).
\label{eq:metricv}
\]
The right-hand sides certainly agree with the standard diffeomorphism in general relativity for a scalar and a metric.
It should be stressed that this result can be obtained, only when the correction $h(\beta,g^{\mu\nu},n)$ with the parameters 
satisfying \eq{eq:condz} is included in the equation of motion as in \eq{eq:eomfinal}.

Now, let us briefly discuss the inclusion of the terms corresponding to those parameterized 
by a shift vector in the Hamiltonian of the ADM formalism of general relativity. 
The last term of the EOM of CTM in \eq{eq:ctmeom} represents an arbitrary infinitesimal 
SO$({\cal N})$ transformation.
As discussed in Section~\ref{sec:gauge}, it contains the diffeomorphism and the spin-three gauge transformation 
in the present context. However, since the latter is used to maintain the gauge-fixing condition $\beta^{\mu\nu\rho}=0$, 
only the diffeomorphism can be set arbitrary.
The diffeomorphism transformation \eq{eq:diffeo} and the identification \eq{eq:geqbeta2} imply that 
$\beta$ and $g^{\mu\nu}$ are transformed in the standard way of general relativity. 
Thus, implementing the following replacement in \eq{eq:eomfinal},
\[
\begin{split}
\dot \beta&\rightarrow  \dot\beta+n^\mu \beta_{,\mu}, \\
\dot g^{\mu\nu}&\rightarrow \dot g^{\mu\nu}-2n^{(\mu,\nu)},
\end{split}
\]
where $n^\mu$ is a newly introduced shift vector, 
one obtains the EOM with the shift vector.

\section{Deletion of the derivatives of the lapse function}
\label{sec:reparametrization}
The equation of motion (EOM) \eq{eq:eomfinal} contains some terms with the derivatives of $n$.
As discussed below \eq{eq:explicitbetas},
this is an obstacle for a general relativistic interpretation of the EOM of CTM.
In this section, we will show that, by redefining the fields $\beta,g^{\mu\nu}$ with some 
derivative corrections, one can actually delete all the terms with the derivatives of $n$ from the EOM.  

The reparameterization of the fields we consider is given by adding some correction terms with second order of derivatives:
\[
\begin{split}
\beta&\rightarrow \beta+x_1\, \frac{\beta_{,\mu}\beta^{,\mu}}{\beta^3}+x_2 \, \frac{\beta_{,\mu}^{,\mu}}{\beta^2}
+x_7\,\frac{R}{\beta}, \\
g^{\mu\nu}&\rightarrow g^{\mu\nu} +x_3\, \frac{\beta^{,\mu}\beta^{,\nu}}{\beta^4}+x_4\, \frac{\beta^{,\mu\nu}}{\beta^3}
+x_5\, \frac{g^{\mu\nu} \beta^{,\rho}\beta_{,\rho}}{\beta^4}+x_6\, \frac{g^{\mu\nu}\beta^{,\rho}_{,\rho}}{\beta^3}
+x_8\, \frac{R^{\mu\nu}}{\beta^2}+x_9\,\frac{g^{\mu\nu} R}{\beta^2},
\label{eq:replacement}
\end{split}
\]
where $x_i$'s are parameters. Note that, since the reparameterization is covariant, the algebraic consistency 
between the commutation of time evolutions and the diffeomorphism obtained so far should be 
unaltered\footnote{For sure, we have checked it through explicit computations.}.

There exist two kinds of effects from this reparameterization.
The first one is on the right-hand side of \eq{eq:eomfinal}. Since the corrections are in the second order of derivatives, 
the reparameterization is effective only on the zeroth order term, 
and causes some shifts of the coefficients of the non-derivative terms of $n$.
On the other hand, the reparameterization affects the left-hand side more importantly for our purpose.
$\dot \beta$ will be replaced by
\[
\begin{split}
\dot \beta \rightarrow & \dot \beta+x_1 \left( - \frac{3\dot \beta \beta_{,\mu}\beta^{,\mu}}{\beta^4}
+\frac{2\dot\beta_{,\mu}\beta^{,\mu}+\dot g^{\mu\nu}\beta_{,\mu}\beta_{,\nu}}{\beta^3}\right)\\
&\ \ \ +
x_2 \left( -\frac{2 \dot \beta \beta_{,\mu}^{,\mu}}{\beta^3}+\frac{\dot \beta_{,\mu}^{,\mu}+\dot g^{\mu\nu}\beta_{,\mu\nu}
-g^{\mu\nu}\dot\Gamma_{\mu\nu}^\rho \beta_{,\rho}}{\beta^2} \right)
+x_7\, \left( -\frac{R\dot \beta}{\beta^2}+\frac{\dot g^{\mu\nu}R_{\mu\nu}+g^{\mu\nu}\dot R_{\mu\nu}}{\beta} \right).
\end{split}
\label{eq:changedotbeta}
\] 
To evaluate the correction terms in \eq{eq:changedotbeta} up to the second order of derivatives, 
we can put the zeroth order expressions of the time-derivative of the fields, 
i.e., the first equation of \eq{eq:zeroth}, \eq{eq:dotg}, \eq{eq:dotgamma}, and \eq{eq:dotR}, 
into them. The things are similar for the correction terms in the replacement of $g^{\mu\nu}$ in \eq{eq:replacement}. 
Then, because the zeroth order expressions contain $n$, there emerge a number of terms which contain
the derivatives of $n$. In fact, we can delete all the derivative terms of $n$ in the EOM 
by appropriately choosing the $x_i$'s.
The condition for the deletion is expressed by six equations, which are explicitly given in Appendix~\ref{app:deleten}. 
Solving the equations for $x_1,\cdots, x_6$, and putting the solutions into the EOM, 
we obtain
\[
\begin{split}
\frac{1}{n}\dot \beta&=
-\frac{3 (-6 + d) \beta^2 }{2 + d}+\frac{(1 + (6 - 9 d) x_7) R}{2 + d}
-\frac{16 (-1 + d) (-1 + (-6 + 9 d) x_7)  
\beta_{,\mu}^{,\mu}}{(-6 + d) (2 + d) \beta}\\
&+\frac{2 (-8 (11 + 84 x_7) + d^3 (-1 + 360 x_7) + 4 d (43 + 480 x_7) - 
     2 d^2 (19 + 804 x_7))   \beta_{,\mu}\beta^{,\mu}}{(-6 + d)^2 (2 + d) \beta^2}\\
&+{\cal O}(\nabla^4),\\
\frac{1}{n} \dot g^{\mu\nu}&=\frac{48 \beta  g^{\mu\nu}  }{2 + d}
+\frac{2  (1 + 24 x_7 - 3 (2 + d) x_9)     g^{\mu\nu}   R}{(2 + d) \beta}
-\frac{2  (1 + 3 x_8)  R^{\mu\nu}}{\beta}\\
&+\frac{A_1 g^{\mu\nu}  \beta_{,\rho}\beta^{,\rho}}{(-6 + d)^2 (2 + d) \beta^3}
+\frac{16  (48 + 84 x_8 + 3 d^2 x_8 - 8 d (1 + 6 x_8))  \beta^{,\mu}\beta^{,\nu}}{(-6 + d)^2 \beta^3}\\
&-\frac{16  (4 + 48 x_7 + 6 x_8 - 12 x_9 + 6 d^2 x_9 + 
    d (-1 - 48 x_7 + 3 x_8 + 6 x_9))  g^{\mu\nu}  
\beta_{,\rho}^{,\rho}}{(-6 + d) (2 + d) \beta^2}\\
&-\frac{8 (-6 + d - 12 x_8 + 6 d x_8)  \beta^{,\mu\nu} }{(-6 + d) \beta^2}+{\cal O}(\nabla^4),
\end{split}
\label{eq:eomCTM}
\]
where 
\[
\begin{split}
A_1&=16 (2 d (13 + 456 x_7 - 6 x_8 - 72 x_9) - 4 (20 + 168 x_7 + 33 x_8 - 42 x_9) + 30 d^3 x_9\\
&\hspace{3cm}- 3 d^2 (1 + 80 x_7 - 9 x_8 + 18 x_9)).
\end{split}
\label{eq:A1}
\]
Interestingly, the EOM does not depend on the two-dimensional ambiguity of the solutions of $z_i$'s to \eq{eq:condz},
and is parameterized solely by $x_{7,8,9}$.
In the following section, we will identify \eq{eq:eomCTM} with 
the EOM of general relativity coupled with a scalar field 
based on the Hamilton-Jacobi approach. 

In the EOM \eq{eq:eomCTM}, one can see that 
the scale transformation \eq{eq:scaletrans} is realized as 
\[
\begin{split}
t&\rightarrow L t,\ x^{\mu}\rightarrow L x^\mu, \\
\beta&\rightarrow \frac{\beta}{L},
\end{split}
\label{eq:scaletranseom}
\]
while $n,g^{\mu\nu}$ are invariant.

\section{Hamilton-Jacobi equation of general relativity coupled with a scalar field}
\label{sec:conttheory}
In this section, starting with an action of general relativity coupled with a scalar field, 
and employing the Hamilton-Jacobi approach, 
we identify the equations of motion (EOM) of this gravitational system with the EOM \eq{eq:eomCTM}  of CTM.

It is an easy task to guess a possible form of the action for the purpose:
\[
S = \int_{\mathcal{M}} d^{d+1}x \sqrt{- G} \left( 2 R^{(d+1)} -\frac{A}{2} G^{ij} \partial_i \phi \partial_i \phi  -\Lambda e^{2 B \phi} \right),
\label{eq:contaction}
\]
where $G_{ij}$ denotes the $(d+1)$-dimensional metric with $i,j=0,1,2,\cdots,d$; 
$R^{(d+1)}$ is the $(d+1)$-dimensional Ricci scalar; $\phi$ is a real scalar field; 
$A,B,\Lambda$ are real parameters. 
The scalar field $\phi$ is assumed to be related to the CTM field $\beta$ through $\beta=e^{B\phi}$. 
This action would be considered to be an effective action valid up to the second order of derivatives. 
The classical EOM derived from \eq{eq:contaction} respects
the dilatational symmetry \eq{eq:scaletranseom}, because $S$ is transformed homogeneously 
by the transformation as $S\rightarrow L^{d-1} S$. 

Considering that the $(d+1)$-dimensional Lorentzian manifold $\mathcal{M}$ is globally hyperbolic, 
we use the following diffeomorphism,
\[
\varphi : \ \Sigma \times \mathbb{R} \to \mathcal{M}, 
\label{eq:diff}
\] 
where $\Sigma$ is a $d$-dimensional spatial hypersurface, 
to obtain the ADM metric as a pull-back, $\varphi^*G$:
\[
ds^2 = - N^2 dt^2 + g_{\mu \nu} (dx^{\mu} + N^{\mu}dt )(dx^{\nu} + N^{\nu}dt),
\]
where $N$, $N^{\mu}$ and $g_{\mu \nu}$ are the lapse function, the shift vector and the $d$-dimensional metric on $\Sigma$ with $\mu, \nu = 1,2, \cdots, d$.  
Hereafter we will turn off the shift vector, i.e., $N^{\mu}=0$ for simplicity. 
The terms associated with the non-zero shift vector can be recovered considering the time-dependent spatial diffeomorphism.

By the diffeomorphism (\ref{eq:diff}), the action (\ref{eq:contaction}) becomes
\[
S= \int dt\ (K - V),
\]
where $K$ is the kinetic term, 
\[
K= \int_{\Sigma_t} d^dx\ 
\left( 
\frac{1}{2} \mathcal{G}^{\mu\nu,\rho\sigma} \dot g_{\mu\nu} \dot g_{\rho\sigma} 
+\frac{1}{2} \mathcal{G}^{\phi,\phi}\dot \phi \dot \phi
\right)
\]
with 
\[
\begin{split}
\mathcal{G}^{\mu\nu,\rho\sigma}&=\frac{\sqrt{g}}{N} \left( \frac{1}{2} (g^{\mu\rho}g^{\nu\sigma}+g^{\mu\sigma}g^{\nu\rho})
-g^{\mu\nu}g^{\rho\sigma}\right),\\
\mathcal{G}^{\phi,\phi}&=\frac{A\sqrt{g}}{N},
\label{eq:quadratic}
\end{split}
\]
and $V$ is the potential term, 
\[
V 
= \int_{\Sigma_t} d^d x\ N \sqrt{g} \left( \Lambda e^{2B \phi} -2 R+\frac{A}{2} (\nabla \phi)^2 \right),
\label{eq:potential}
\]
in which $(\nabla \phi)^2 := g^{\mu \nu} \nabla_{\mu} \phi \nabla_{\nu} \phi$ with $\nabla_{\mu}$ being the covariant derivative 
associated with the metric $g_{\mu \nu}$.

To employ the Hamilton-Jacobi formalism, 
let us consider the following Hamilton's principal functional:  
\[
W =\int_{\Sigma_t} d^d x \sqrt{g} \left( \lambda e^{B\phi} -e^{-B\phi} \left( c_1 R+c_2 (\nabla \phi)^2 \right)\right)+{\cal O}({\nabla^4}),
\label{eq:W}
\]
where $c_1,c_2,\lambda$ are real parameters. 
$W$ is considered to be expressed as 
perturbative expansions in spatial derivatives up to the second order. 
The potential in \eq{eq:potential} and 
$W$ in \eq{eq:W} must be related by the following Hamilton-Jacobi equation: 
\[
V + \int_{\Sigma_t} d^dx\ 
\frac{1}{2}\left(
\mathcal{G}_{\mu\nu,\rho\sigma} \frac{\delta W}{\delta g_{\mu\nu}} \frac{\delta W}{\delta g_{\rho\sigma}}
+\mathcal{G}_{\phi,\phi} \frac{\delta W}{\delta \phi} \frac{\delta W}{\delta \phi} 
\right) +{\cal O}(\nabla^4)
= 0,
\label{eq:VandW}
\]
where 
\[
\begin{split}
\mathcal{G}_{\mu\nu,\rho\sigma}&=\frac{N}{\sqrt{g}}\left(\frac{1}{2} (g_{\mu\rho}g_{\nu\sigma}+g_{\mu\sigma}g_{\nu\rho})
-\frac{1}{d-1}g_{\mu\nu}g_{\rho\sigma}\right),\\
\mathcal{G}_{\phi,\phi}&=\frac{N}{A\sqrt{g}},
\end{split}
\]
being the inverse to \eq{eq:quadratic}.
Inserting \eq{eq:W} into \eq{eq:VandW}, 
we obtain
\[
V = 
\int_{\Sigma_t} d^dx\ 
\sqrt{g}N \,
\frac{1}{2}\left[
\frac{1}{d-1} 
\left( \frac{\lambda^2de^{2 B\phi}}{4} +\lambda H \right)
+\frac{1}{A} \left(2 B \lambda F -B^2 \lambda^2 e^{2 B \phi} \right)
\right] +{\cal O}(\nabla^4),
\label{eq:V}
\]
where
\[
\begin{split}
H&=\frac{2-d}{2}\left( c_1 R+c_2 (\nabla \phi)^2\right)+(d-1)c_1 \left(B \nabla^2 \phi+B^2 (\nabla\phi)^2\right),\\
F&=-B\left(c_1 R -c_2 (\nabla \phi)^2\right)-2 c_2 \nabla^2 \phi.
\end{split}
\label{eq:HandS}
\]
Comparing \eq{eq:V} with \eq{eq:potential}, we obtain some conditions for the parameters of $W$ as 
\[
\begin{split}
\Lambda&=\frac{\lambda^2(-4 B^2 (-1 + d) + A d)}{8 A (-1 + d)},\\
-2&=-\frac{c_1 \lambda(A (-2 + d) + 4 B^2 (-1 + d))}{4 A (-1 + d))},\\
\frac{A}{2}&=\frac{\lambda(A (-c_2 (-2 + d) + 2 B^2 c_1 (-1 + d)) + 
   4 B^2 c_2 (-1 + d))}{4 A (-1 + d)},\\
0&=B \lambda (A c_1 + 4 c_2).
\end{split}
\label{eq:relpara}
\]
Here, the first equation comes from the comparison of the potential term, the second the curvature, and the third the scalar 
kinetic term. The last equation comes from the absence of $\nabla^2 \phi$ term in the potential. 

The flow equations derived from $W$ is given by\footnote{
The flow equations for $\phi$ and $g_{\mu \nu}$  
are originated with Hamilton's equations for $\phi$ and $g_{\mu \nu}$ 
with the replacement of conjugate momenta by $\frac{\delta W}{\delta \phi}$ and $\frac{\delta W}{\delta g_{\mu \nu}}$, 
respectively.   
} 
\[
\begin{split}
&\frac{1}{N} \dot \phi=\mathcal{G}_{\phi,\phi} \frac{\delta W}{\delta \phi} 
=\frac{1}{A} \left(B \lambda e^{B\phi}-e^{-B\phi} F \right) + {\cal O}(\nabla^4), \\
&\frac{1}{N}\dot g_{\mu\nu}=  \mathcal{G}_{\mu\nu,\rho\sigma}\frac{\delta W}{\delta g_{\rho\sigma}}
=\frac{\lambda e^{B\phi}}{2(1-d)}g_{\mu\nu}+e^{-B\phi}\left(H_{\mu\nu}+\frac{1}{1-d} g_{\mu\nu} H\right)
+{\cal O}(\nabla^4), 
\end{split}
\label{eq:eommetbare}
\]
where
\[
\begin{split}
H_{\mu\nu}=&c_1\left( R_{\mu\nu}-\frac{1}{2}g_{\mu\nu}R+B \left( \nabla_\mu\nabla_{\nu}\phi-g_{\mu\nu}\nabla^2 \phi \right)
-B^2 \left(\nabla_\mu\phi \nabla_\nu\phi-g_{\mu\nu} (\nabla \phi)^2 \right)\right) \\
&+c_2 \left(\nabla_\mu\phi \nabla_\nu\phi-\frac{1}{2}g_{\mu\nu} (\nabla \phi)^2 \right).
\end{split}
\]
There is a relation, $H=g^{\mu\nu}H_{\mu\nu}$. 

To compare \eq{eq:eommetbare} with the EOM \eq{eq:eomCTM} from CTM, 
let us perform 
a change of the variable, $\beta=\exp[B \phi]$. Taking into account that 
$\dot g^{\mu\nu}=-g^{\mu\mu'}g^{\nu\nu'}\dot g_{\mu'\nu'}$, 
the EOM \eq{eq:eommetbare} can be rewritten as 
\[
\begin{split}
\frac{1}{c_3 n}\dot \beta=&\frac{B^2 \lambda \beta^2}{A}+\frac{B^2 c_1 R}{A}+\frac{2 c_2 \beta_{,\mu}^{,\mu}}{A \beta}
-\frac{3 c_2 \beta_{,\mu}\beta^{,\mu}}{A \beta^2}+ {\cal O}(\nabla^4) ,\\
\frac{1}{c_3 n} \dot g^{\mu\nu}=&\frac{\lambda \beta  g^{\mu\nu}}{2 (-1 + d)}
-\frac{c_1 R^{\mu\nu}}{\beta}
-\frac{c_1  \beta^{,\mu\nu}}{\beta^2}
+\frac{(2 B^2 c_1 - c_2) \beta^{,\mu}\beta^{,\nu}}{B^2 \beta^3} \\
&+g^{\mu\nu}\left( \frac{c_1 R}{2 (-1 + d) \beta}
+\frac{c_2 \beta_{,\rho} \beta^{,\rho}}{2 B^2 (-1 + d) \beta^3}
\right)+ {\cal O}(\nabla^4),
\end{split}
\label{eq:eomcont}
\]
where we have introduced possible difference of normalizations between the lapse functions 
of CTM and general relativity as $N=c_3\, n$ with a constant $c_3$. 
We want to find the values of the parameters which make \eq{eq:eomcont} coincident with \eq{eq:eomCTM}.
The number of parameters is smaller than that of the equations to be satisfied (i.e., an overdetermined set of equations), 
but we can solve the coincidence condition by the following values:
\[
\begin{split}
\lambda c_3 &=\frac{96(-1+d)}{2+d}, \\
B^2&=\frac{A(6-d)}{32(d-1)},\\
c_1 c_3&=\frac{8(2+d)}{-10+7d},\\
c_2 c_3 &=-\frac{2 A (2 + d)}{-10 + 7 d}, \\
x_7&=\frac{16 - 88 d + 26 d^2 + d^3}{12 (-20 + 64 d - 65 d^2 + 21 d^3)},\\
x_8&=\frac{6 - d}{-10 + 7 d}, \\
x_9&=\frac{14 - 67 d + 17 d^2}{-60 + 192 d - 195 d^2 + 63 d^3}.
\end{split}
\label{eq:detpara}
\]
The details of the derivation of the solution are given in Appendix~\ref{app:solution}.
The parameter $c_3$ can be determined by the second (or equivalently the third) 
equation of \eq{eq:relpara} by putting \eq{eq:detpara}:
\[
c_3^2=12.
\label{eq:valuec3}
\] 
This rather strange value actually normalizes the overall factor
in the algebraic relation \eq{eq:metricv} of the CTM to the natural value in GR.
It can be checked that the third and fourth equations
of \eq{eq:relpara} are also satisfied by \eq{eq:detpara} and \eq{eq:valuec3}. 
From the first equation of \eq{eq:relpara}, \eq{eq:detpara} and \eq{eq:valuec3},  we obtain
\[
\Lambda=\frac{36 (d-1)(3d-2)}{(d+2)^2}.
\label{eq:valuelambda}
\]
The above solution is unique except for the rather obvious ambiguities of the signs of $B$ and $c_3$.
These signs are physically irrelevant, because the sign of $B$ can be absorbed by that of $\phi$, and 
that of $c_3$ just determines the overall sign of $W$ (or can be absorbed in $n$).

If we require the positivity of the potential energy from the spatial derivative term of $\phi$, $A>0$ is required. Then, 
the second equation of \eq{eq:detpara} implies that the dimension must be in the range 
$2\leq d\leq 6$ (The $d=1$ case is excluded from the beginning in the Hamilton formalism,  
as can be seen at the beginning of this section.).
In this range, \eq{eq:valuelambda} is positive, and one can normalize the value of $\Lambda$ 
by rescaling the space-time coordinates as $(t,x^\mu)\rightarrow L (t,x^\mu)$ with $L=1/\sqrt{\Lambda}$ and 
dropping an overall factor of the action.
We can also rescale the scalar field as $\phi \rightarrow \hbox{sign(}B\hbox{)}\phi/\sqrt{A}$.
Then, the action describing CTM is uniquely determined, 
for a globally hyperbolic $\mathcal{M}$, 
to be 
\[
S_{CTM}=\int_{\mathcal{M}} d^{d+1} x\,\sqrt{-G} \left( 
2 R-\frac{1}{2} G^{ij} \partial_{i} \phi \partial_{j} \phi 
- e^{\sqrt{\frac{6-d}{8(d-1)}}\phi}
\right),
\label{eq:CTMaction}
\]
which is valid in $2\leq d \leq 6$. Thus, 
the system has a critical dimension $d=6$, over which it becomes unstable due to the 
wrong sign of the scalar kinetic term. 
At the critical dimension,  the scalar is a massless field with no non-derivative couplings.

\section{Time evolution of the scale factor}
\label{sec:mini}
The coupled system of gravity and a scalar field described by the action \eq{eq:CTMaction} 
has been discussed in the context of models of dark energy 
(See \cite{Copeland:2006wr} for a comprehensive review.).
The exponential potential in \eq{eq:CTMaction} of the scalar field is known to lead to a power-law 
behavior (or an exponential behavior in the critical case) of the scale factor. 
Let us see this in our case, analyzing \eq{eq:eomCTM}.

Discarding the spatial derivative terms of \eq{eq:eomCTM} and putting $n=1$, the equation of motion is given by
\[
&\dot \beta=d_1 \beta^2,
\label{eq:minibeta} \\
&\dot g^{\mu\nu}=d_2 \beta g^{\mu\nu},
\label{eq:minig}
\]
where 
\[
d_1=\frac{3(6-d)}{d+2},\ d_2=\frac{48}{d+2}.
\]
Substituting \eq{eq:minig} with an ansatz $g^{\mu\nu}=a(t)^{-2} \delta^{\mu\nu}$ 
with a scale factor $a(t)$, we obtain
\[
\frac{2\dot a}{a}=-d_2 \beta. 
\label{eq:minia}
\]
Then, for $d_1\neq 0\ (\hbox{i.e., }d\neq 6)$, the solution to \eq{eq:minibeta} and \eq{eq:minia} is obtained as
\[
\begin{split}
&\beta=\frac{1}{d_1(t_0-t)},\\
&a=a_0 (t_0-t)^{\frac{d_2}{2d_1}},
\end{split}
\]
where $t_0$ and $a_0$ are integration constants.

When $d=6$, $d_1$ vanishes.  In this case, $\beta$ is given by a constant, say $\beta_0$. Then, 
\eq{eq:minia} gives
\[
a=a_0 \exp \left[ -\frac{d_2 \beta_0}{2} t \right].
\]
Thus, we see that, in the critical case $d=6$, the solution is given by de Sitter spacetime.

As is well known, de Sitter spacetime has the invariance of a conformal symmetry $SO(d+1,1)$. 
In statistical physics, the appearance of a conformal symmetry is the sign that a system is on a critical point. 
This suggests that CTM at $d=6$ is on a critical point in some sense. In fact, as shown in Section~\ref{sec:conttheory}, 
for the reality of $B$, the sign of the kinetic term of the scalar field must change its sign at $d=6$.
In $d>6$, it gets the wrong sign, and the the scalar field becomes unstable in the direction 
of larger spatial fluctuations.
This means that $d=6$ can be thought of as a phase transition point 
between a stable phase at $d<6$ and another phase at $d>6$.
Considering the instability in the direction of larger spatial fluctuations, 
the latter phase probably contradicts our assumption of a continuous
space. The understanding of the phase transition should be pursued further.

\section{Summary and future prospects}
\label{sec:summary}
In this paper, we have analyzed the equation of motion (EOM) of the canonical tensor model (CTM)
in a formal continuum limit by employing a derivative expansion of its tensor up to the fourth order. 
We have shown that, up to the order, the EOM of CTM in the continuum limit agrees with 
that of a coupled system of gravity and a scalar field obtained in the framework of the Hamilton-Jacobi methodology. 
The action of the gravitational system is composed of the curvature term, the scalar field kinetic term
and an exponential potential of the scalar field. The system is classically invariant under a dilatational transformation. 
The action is physically valid in the range of the spatial dimensions, $2\leq d \leq 6$,
and, in $d>6$, the system is unstable due to the wrong sign of the kinetic term of the scalar field.
At the critical case $d=6$, de Sitter spacetime is a solution to the EOM, while, in $2\leq d < 6$,
the time evolution of the scale factor of a flat space has a power-law behavior.
 
The most significant achievement of this paper is to have concretely shown that CTM indeed derives a 
general relativistic system in a formal continuum limit. This was conjectured in our previous paper \cite{Sasakura:2015pxa}
from the observation that
the constraint algebra of CTM in the continuum limit agrees with that of the ADM formalism, 
but no concrete correspondences were given. 
On the other hand, in this paper, we have obtained the one-to-one correspondence of the fields 
between CTM in the continuum limit up to the fourth order 
and the gravitational system so that the two systems have a common EOM.
The action of the corresponding gravitational system has also been obtained.

An interesting question arising from our result is what is the meaning of the criticality at $d=6$.
The existence of de Sitter spacetime solution implies that the system has a conformal symmetry on this background
in the dimension.
On the other hand, in  our previous papers \cite{Sasakura:2015xxa,Sasakura:2014zwa,Sasakura:2014yoa}, it was shown 
that the Hamiltonian vector flows of CTM can be regarded as RG flows of statistical systems on random networks.
These two aspects of CTM suggest that a statistical system at criticality described by a 
six-dimensional conformal field theory is associated to CTM \cite{Strominger:2001pn,Strominger:2001gp}.
It would be interesting to identify the conformal field theory in a concrete manner.

Another interesting direction of study would be to extend the derivative expansion to higher orders,
which includes higher spin fields than two with higher spin gauge symmetries. 
There are general interests in pursuing higher spin gauge theories (See \cite{Giombi:2016ejx} for a recent review.).
Since our approach has a significant difference from the other ones in the sense 
that we take a formal continuum limit of a consistent 
discretized theory in the canonical formalism, we would expect that our model may shed some new lights on the subject. 
For that purpose, it would be necessary to set up a new efficient methodology for the analysis instead of relying on machine 
powers as in this paper.

The EOM of CTM is a set of first-order differential equations in time, and has been related  
to a gravitational system through the Hamilton-Jacobi equation.
While the gravitational system contains the phenomena of second-order differential equations like wave propagations,
it is not clear how to realize such phenomena in the framework of CTM. 
It would be interesting to improve the canonical formalism of the tensor model in that direction.

\vspace{1cm}
\centerline{\bf Acknowledgements} 
The work of N.S. is supported in part by JSPS KAKENHI Grant Number 15K05050. 
The work of Y.S. is funded under CUniverse research promotion project 
by Chulalongkorn University (grant reference CUAASC).
N.S. would like to thank the great hospitality and the stimulating discussions with the members
of Department of Physics of Chulalongkorn University, while he stayed there and part of this work was done. 
Y.S. would like to thank the wonderful members in Nagoya University, Japan, 
where part of this work was done, for the kind hospitality and fruitful discussions. 
Y.S. would like to appreciate Jan Ambj\o rn for discussions about the dimension in the formal continuum limit.   

\appendix

\section{Characterization of a symmetric rank-three tensor}
\label{app:pf3}
A totally symmetric rank-three tensor $P_{abc}$ can be fully characterized by 
the values of $P_{abc}\phi_a\phi_b\phi_c$ for arbitrary vector $\phi$.
To prove this, let us show that $P_{abc}\phi_a^1\phi_b^2\phi_c^3$ for arbitrary three vectors $\phi^{1,2,3}$ 
can be computed from $P_{abc}\phi_a\phi_b\phi_c$ with some $\phi$'s.

Let us define
\[
\begin{split}
\tilde \phi^0&:=\phi^1+\phi^2+\phi^3, \\
\tilde \phi^1&:=-\phi^1+\phi^2+\phi^3, \\
\tilde \phi^2&:=\phi^1-\phi^2+\phi^3, \\
\tilde \phi^3&:=\phi^1+\phi^2-\phi^3.
\end{split}
\]
Then, one finds
\[
24 P_{abc} \phi^1_a  \phi^2_b \phi^3_c=P_{abc}\tilde \phi^0_a \tilde \phi^0_b \tilde \phi^0_c
-P_{abc}\tilde\phi^1_a \tilde\phi^1_b \tilde\phi^1_c
-P_{abc}\tilde\phi^2_a \tilde\phi^2_b \tilde\phi^2_c
-P_{abc}\tilde\phi^3_a \tilde\phi^3_b \tilde\phi^3_c.
\] 

\section{Independent fields up to the fourth order}
\label{app:fourth}
In this section, we will explain the reason why we can take the independent fields as in \eq{eq:pf3}
up to the fourth order. 

Let us first discuss the necessity of the symmetrization of  the covariant derivatives of $f$. 
As an example, let us consider the third covariant derivative of $f$:
\[
\nabla_\mu\nabla_\nu \nabla_\rho f.
\]
From the definition of the covariant derivative,
$\nabla_\mu\nabla_\nu \nabla_\rho f$ is symmetric between $\nu$ and $\rho$, but generally not between $\mu$ and $\nu$. 
The anti-symmetric part between $\mu$ and $\nu$ is given by 
\[
\nabla_\mu\nabla_\nu \nabla_\rho f -\nabla_\nu\nabla_\mu \nabla_\rho f=R_{\mu\nu\rho}{}^\sigma \nabla_{\sigma}f.
\]
Therefore, the anti-symmetric part can be absorbed into the first derivative term. 
Similar things occur also in the other derivative terms.
Thus, 
to secure the uniqueness of the representation of $P$ 
in terms of the fields, it is necessary to symmetrize the covariant derivatives of $f$.   
Correspondingly, we have to assume the index symmetries of the $\beta$'s explained below \eq{eq:pf3} for the 
unique characterization.

Next, let us discuss the absence of some fields in the expansion \eq{eq:pf3}. 
Generally, $Pf^3$ may contain a term,
\[
\int d^d x \, \beta^\mu f^2 f_{,\mu}.
\] 
However, by performing partial integrations, we obtain
\[
\int d^d x \, \beta^\mu f^2 f_{,\mu}=\frac{1}{3}\int d^d x \, \beta^\mu (f^3)_{,\mu}=-\frac{1}{3}\int d^d x \, \beta^\mu_{,\mu} f^3,
\]
where the test function $f$ is assumed to have a compact support.
Therefore,  $\beta^\mu$ is not independent and can be absorbed into $\beta$. 
The same argument can be extended to the other possible terms. The most non-trivial term would be
\[
\int d^d x\, \beta^{\mu\nu\rho\sigma} f^2 f_{,\mu\nu\rho\sigma},
\]
where $\beta^{\mu\nu\rho\sigma}$ is assumed to be symmetric as explained above. By partial integrations, we obtain 
\[
\int d^d x  \left( \frac{1}{9} \beta^{\mu\nu\rho\sigma}_{,\mu\nu\rho\sigma} f^3 -2  \beta^{\mu\nu\rho\sigma}_{,\mu\nu}
f^2 f_{,\rho\sigma} -\frac{8}{3} \beta^{\mu\nu\rho\sigma}_{,\mu} f^2 f_{,\nu\rho\sigma}+2\beta^{\mu\nu\rho\sigma}
f f_{,\mu\nu}f_{,\rho\sigma}- \beta^{\mu\nu\rho\sigma}
f^2 f_{,\mu\nu\rho\sigma}\right)=0.
\label{eq:ambiguity}
\] 
Therefore,  $\beta^{\mu\nu\rho\sigma}$ can be absorbed into the fields existing in \eq{eq:pf3}. 

\section{Derivative expansion and associativity}
\label{app:derivative}
In this paper, we consider a formal continuum limit, and express $P$ in terms of the derivative expansion \eq{eq:pf3}
with a termination at a certain order.
This approximation with a cut-off in the number of derivatives 
can be physically validated by assuming that the scale of our physical interest is much larger than that of 
the fundamental fuzziness of the space.
However, from theoretical view points, this approximation must be checked with much care, because 
any approximations generally contain 
the potential risk of destroying the essential part of the framework of CTM, namely, the algebraic closure of the constraints.
This is directly related to the covariance of the spacetime interpretation of CTM 
in the continuum limit, and therefore, its slightest violations would lead to pathological behaviors of dynamics
ruining a consistent picture.  
The procedure we take in the approximation is that, for each term in the derivative expansion,
we sum up the numbers of the derivatives of the fields, the background metric, and the test functions, 
and we neglect the term if the sum exceeds a certain number, which is four in this paper.
In this section, we will explain the algebraic consistency of our procedure.

In the equation of motion \eq{eq:ctmeom} of CTM, the closure of the algebraic structure of the constraints
appears in the commutation of two successive infinitesimal time evolutions:
the commutation is given by an infinitesimal SO$({\cal N})$ transformation. 
More precisely, for the infinitesimal time evolutions represented by 
$n_1$ and $n_2$ (the second term in \eq{eq:ctmeom} is ignored for brevity), respectively, one can show that
\[
\left(\delta_{n_1}\delta_{n_2}-\delta_{n_2} \delta_{n_1}\right) P_{abc}\propto 
\sum_{\sigma}\  [\tilde n_1, \tilde n_2]_{\sigma_a d}P_{d \sigma_b \sigma_c},
\label{eq:comtimegen}
\] 
where $\tilde n_{ab}:= n_c P_{cab}$. The right-hand side is indeed an infinitesimal SO$({\cal N})$
transformation of $P$ with a gauge parameter $[\tilde n_1, \tilde n_2]$.

In the derivation of \eq{eq:comtimegen}, an implicit assumption is the following simple property of the index contraction:
\[
P_{abc}P_{cde}P_{efg}=(P_{abc}P_{cde})P_{efg}=P_{abc}(P_{cde}P_{efg}),
\label{eq:tensorassos}
\] 
where the parentheses of the last two expressions represent which of the two indices, $c$ or $e$, 
are first summed over.  This associativity of the tensor manipulation is trivial for a finite ${\cal N}$,
but it would not be generally true in an infinite case as in a continuum limit especially with a cut-off.  
We need to check if our procedure of the approximation mentioned above is consistent with the associativity.

By contracting all the other irrelevant indices than $c$ or $e$  in \eq{eq:tensorassos} with some vectors, 
the associativity is reduced to the question whether $(A_a B_{ab}) C_b=A_a (B_{ab} C_b)$.  
In our case, an index contraction of tensors is represented by the covariant derivatives and the integrations
as in \eq{eq:pf3}.
Thus, let us consider 
\[
\begin{split}
A[f]&:=\int dx \sum_{0\leq p+q \leq m} \alpha^p_q \partial^q f ,\\
B[f,g]&:=\int dx \sum_{0\leq p+q+r \leq m} \beta^p_{qr} (\partial^q f) (\partial^r g),\\
C[f]&:=\int dx \sum_{0\leq p+q \leq m} \gamma^p_q \partial^q f ,
\end{split}
\]
where $f,g$ are test functions. Here, for simplicity, we take the notation of the one-dimensional case with partial derivatives,
but the discussions should be extendible to higher dimensions with covariant derivatives.
The upper indices of $\alpha,\beta,\gamma$ represent the numbers of derivatives intrinsically contained in 
$\alpha,\beta,\gamma$, respectively, and $m$ represents the cut-off on the order of derivatives we consider.
Let us remind that, in our procedure, 
the number of derivatives is counted by summing the numbers of the derivatives of all the components, namely,
the fields, the background metric and the test functions.  

The associativity to be shown is that $B[A,C]$ does not depend on whether we first compute $B[A,\cdot ]$ or $B[\cdot ,C]$. 
In the former case, we obtain (contracted with a test function $g$)
\[
\begin{split}
B[A,g]&=B\left[\frac{\delta }{\delta f}A[f],g\right] \\
&=\int dx \sum_{0\leq p+q+r +p'+q' \leq m} \beta^p_{qr} (\partial^{q+q'} (-)^{q'} \alpha^{p'}_{q'}) (\partial^r g),
\end{split}
\label{eq:BAg}
\]
where we have defined the component $A_a$ from $A[f]$ with a functional derivative in the same way as \eq{eq:pff}.
Then, by inserting $g=\frac{\delta }{\delta f}C[f]$, we obtain
\[
B[A,C]=\int dx \sum_{0\leq p+q+r +p'+q'+p''+q'' \leq m} 
\beta^p_{qr} (\partial^{q+q'} (-)^{q'} \alpha^{p'}_{q'}) (\partial^{r+q''} (-)^{q''} \gamma^{p''}_{q''}).
\]
The expression obviously does not change, even if we first compute $B[\cdot ,C]$. 
The associativity is proven.

A comment is in order. Instead of computing in the way of \eq{eq:BAg}, one can take a 
different manner as 
\[
\begin{split}
B[A,g]&=A\left[  \frac{\delta}{\delta f} B[f,g]\right] \\
&=\int dx \sum_{0\leq p'+q'+p+q+r \leq m} \alpha^{p'}_{q'} \partial^{q'} ((-)^{q}( \partial^q(\beta^p_{qr} \partial^r g)))
\end{split}
\]
for the same quantity. In fact, by partial integrations\footnote{The test functions are assumed to have compact supports.},
this is equivalent to \eq{eq:BAg}, and the computation is unique.

\section{Computations of EOM} 
\label{app:explicit}
We used a Mathematica package ``xTensor" to perform the tensorial computations in this paper.
To obtain EOM \eq{eq:explicitbetas}, instead of computing $\delta P f^3$ in \eq{eq:delpf3}, 
we have computed $\delta P [f,f]= P[n,P[f,f]]+2 P[f,P[n,f]]$, and have extracted 
$\delta \beta$'s through \eq{eq:pff}. This is because \eq{eq:pff} can be used to 
straightforwardly get $\delta \beta$'s from the coefficients 
of the derivative expansions of $f$ in $\delta P[f,f]$, while the expression of $\delta P f^3$ contains an integration, 
which requires the non-trivial task of 
appropriate arrangement of partial integrations to obtain an expression in the form \eq{eq:canpf3}.
Of course, the two ways give the same result.

In the actual computation using ``xTensor", we considered partial derivatives rather than covariant derivatives.
This substantially reduced the number of terms which appeared in the raw result, 
because of the commutative character of partial derivatives, namely, $\partial_\mu \partial_\nu = \partial_\nu \partial_\mu$,
while  $\nabla_\mu \nabla_\nu \neq \nabla_\nu \nabla_\mu$ in general\footnote{It was not easy for us to make a 
program to take care of this duplication automatically in the case of using covariant derivatives. 
Researchers with better programming skills may directly use covariant derivatives in the computation. But the 
final result should agree with ours.}.    
Then, after getting the final form, we promoted the partial derivatives to the covariant derivatives. 
The possible differences between before and after the promotion are the appearances of curvature tensors 
originating from the non-commutativity of covariant derivatives. 
Within our approximation ignoring certain higher orders of derivatives, 
it is not difficult to realize that relevant terms with Riemann tensors can only emerge from the term $f^2 f_{,\mu\nu\rho\sigma}$.   
Then, it is not difficult to explicitly compute the precise expression of this specific term by using covariant derivatives.
Our result is
\[
\delta P f^3 \supset \left(\frac{4}{3} \beta^{\mu\nu}\beta^{\rho\sigma} +2\beta^{\mu\nu,\rho\sigma}\right)
f^2 \left(f_{,\mu\nu\rho\sigma}-f_{,\mu\rho\nu\sigma}\right).
\]
The content of the latter bracket can be computed as 
\[
\begin{split}
f_{,\mu\nu\rho\sigma}-f_{,\mu\rho\nu\sigma}&=\nabla_\mu \left( R_{\nu\rho\sigma}{}^\delta f_{,\delta}\right) \\
&=\left(\nabla_\mu  R_{\nu\rho\sigma}{}^\delta \right) f_{,\delta}+ R_{\nu\rho\sigma}{}^\delta f_{,\mu\delta}.
\end{split}
\]
After the transformation to the form \eq{eq:canpf3}, the first term contributes to $\delta \beta$ in the order 
of ${\cal O}(\nabla^4)$, and can be ignored. On the other hand, the second term contributes to $\delta \beta^{\mu\nu}$ 
as appearing in \eq{eq:explicitbetas}.

Another similar possibility would exist for the term $f^2f_{,\mu\nu\rho}$. This term may produce 
a curvature term of the form $R_{\mu\nu\rho}{}^\sigma f^2f_{,\sigma}$.
However, after the transformation to the form \eq{eq:canpf3}, this merely generates 
a ${\cal O}(\nabla^4)$ term in $\delta \beta$, and is ignorable. There are some other possibilities from the covariant derivatives 
applied to $\beta$'s, but one can easily find that they vanish due to the symmetric properties of the fields or are 
ignorable up to the order of our consideration.

\section{Deleting the derivatives of the lapse function} 
\label{app:deleten}
By performing the computation explained in the text, we obtain the following six equations for 
the absence of the derivatives of $n$ in the equation of motion:
\[
\begin{split}
n_{,\mu}^{,\mu}\hbox{ in }\dot \beta &:
92 - 72 x_2 + 192 x_7 + d^2 (3 + 6 x_2 - 96 x_7 - 6 z_4) \\
&\hspace{3cm}- 4 d (5 + 6 x_2 + 24 x_7 - 6 z_4) + 72 z_4=0,
\\
\beta^{,\mu}n_{,\mu}\hbox{ in }\dot \beta &:
4 (-2 + 18 x_1 + 60 x_2 - 48 x_7 - 9 z_2) + 
  4 d (10 + 6 x_1 + 12 x_2 + 24 x_7 - 3 z_2) \\
&\hspace{6cm} + d^2 (2 - 6 x_1 - 36 x_2 + 96 x_7 + 3 z_2)= 0, \\
g^{\mu\nu}n_{,\rho}^{,\rho}\hbox{ in }\dot g^{\mu\nu} &:-36 x_6 + 3 d^2 (x_6 - 16 x_9) - 4 d (8 + 3 x_6 + 6 x_8 + 12 x_9 - 12 z_4)\\
&\hspace{6cm} + 
 16 (5 - 3 x_8 + 6 x_9 + 6 z_4)=0,\\
g^{\mu\nu} \beta^{,\rho} n_{,\rho}\hbox{ in }\dot g^{\mu\nu} &: 3 d^2 (x_5 + 6 x_6 - 16 x_9) + 
 4 d (-8 + 3 x_4 - 3 x_5 - 6 x_6 - 6 x_8 - 12 x_9 + 6 z_2)  \\
&\hspace{3cm}+ 4 (4 + 6 x_4 - 9 x_5 - 30 x_6 - 12 x_8 + 24 x_9 + 12 z_2) = 0,\\
n^{,\mu\nu}\hbox{ in }\dot g^{\mu\nu}&:d (-2 + 3 x_4 - 24 x_8) + 6 (2 - 3 x_4 + 8 x_8)=0,\\
n^{,\mu}\beta^{,\nu} \hbox{ in }\dot g^{\mu\nu}&: 8 - 6 x_3 - 20 x_4 + d (-4 + x_3 + 2 x_4 - 8 x_8) + 16 x_8=0.
\end{split}
\]

\section{Derivation of  \eq{eq:detpara} and \eq{eq:valuec3}}
\label{app:solution}
In this appendix, we will show some details of the derivation of \eq{eq:detpara} and \eq{eq:valuec3}.

The condition for \eq{eq:eomcont} to be equal to \eq{eq:eomCTM} is given by the following set of equations:
\[
\frac{B^2 c_3 \lambda}{A}&=-\frac{3 (-6 + d)}{2 + d}, 
\label{eq:app1}\\
\frac{B^2 c_1 c_3}{A}&=\frac{1 + (6 - 9 d) x_7}{2 + d},
\label{eq:app2}\\
\frac{2 c_2 c_3}{A}&=-\frac{16 (-1 + d) (-1 + (-6 + 9 d) x_7)}{(-6 + d) (2 + d)},
\label{eq:app3}\\
-\frac{3 c_2 c_3}{A}&=\frac{2 (-8 (11 + 84 x_7) + d^3 (-1 + 360 x_7) + 4 d (43 + 480 x_7) - 
     2 d^2 (19 + 804 x_7))  }{(-6 + d)^2 (2 + d) },
\label{eq:app4}\\
\frac{c_3 \lambda}{2 (-1 + d)} &=\frac{48}{2 + d},
\label{eq:app5}\\
-c_1 c_3&=-2 (1 + 3 x_8),
\label{eq:app6}\\
-c_1 c_3&= -\frac{8 (-6 + d - 12 x_8 + 6 d x_8)}{-6 + d},
\label{eq:app7}\\
\frac{(2 B^2 c_1 - 
   c_2) c_3}{B^2}&=\frac{16 (48 + 84 x_8 + 3 d^2 x_8 - 8 d (1 + 6 x_8))}{(-6 + d)^2},
\label{eq:app8}\\
\frac{c_1 c_3}{2 (-1 + d)}&=\frac{2 (1 + 24 x_7 - 3 (2 + d) x_9)}{2 + d},
\label{eq:app9}\\
\frac{c_2 c_3}{2 B^2 (-1 + d)}&=\frac{A_1}{(-6 + d)^2 (2 + d)},
\label{eq:app10}\\
0&=-\frac{16 (4 + 48 x_7 + 6 x_8 - 12 x_9 + 6 d^2 x_9 + 
    d (-1 - 48 x_7 + 3 x_8 + 6 x_9))}{(-6 + d) (2 + d)},
\label{eq:app11}
\]
where $A_1$ is given by \eq{eq:A1}. The equations have been obtained by equating the corresponding coefficients between 
\eq{eq:eomcont} and \eq{eq:eomCTM}, and have been ordered in the same order as appearing in \eq{eq:eomcont}, except for 
the last one, which is missing in \eq{eq:eomcont} but exists as the coefficient of the $g^{\mu\nu}\beta^{,\rho}_{,\rho}$ term
in the equation for $\dot g^{\mu\nu}$ in \eq{eq:eomCTM}.

The equations can uniquely be solved in the following manner (up to the obvious sign ambiguity of $B$ and $c_3$,
which is physically irrelevant). 
Up to the obvious overall factor of $c_3$, $\lambda$ can be determined from \eq{eq:app5}. $c_1$ and $x_8$ 
can be determined from the set of equations \eq{eq:app6} and \eq{eq:app7}.
$c_2$ and $x_7$ can be determined from the set of \eq{eq:app3} and \eq{eq:app4}. 
Then, by putting $\lambda$ into \eq{eq:app1}, $B^2$ can be determined. By putting $c_1$ and $x_7$ into \eq{eq:app9},
$x_9$ can be determined. These are the solution written in \eq{eq:detpara}. Then, it can be checked that 
all the remaining equations are satisfied by the solution.
 
There are the other set of equations \eq{eq:relpara}, which comes from the Hamilton-Jacobi equation. By putting the above 
solution into the second and the third equations of \eq{eq:relpara}, 
one obtains the same equation $c_3^2=12$, namely \eq{eq:valuec3} (up to sign). 
The forth equation is satisfied by the above solution. The first equation gives the value of the 
cosmological constant (in a broad sense) \eq{eq:valuelambda}.


\end{document}